\documentclass[preprint]{emulateapj}

\usepackage{color}
\usepackage{natbib,graphicx,amsmath}
\def\cii {[C{\scriptsize II}]}
\def\lsol {L_{\odot}}
\def\msol {M_{\odot}}

\def\Arizona{1}
\def\Texas{2}
\def\Diego{3}
\def\Marseille{4}
\def\Dalhousie{5}
\def\ESOGarching{6}
\def\UFlorida{7}
\def\DAWN{8}
\def\UCL{9}
\def\Flatiron{10}
\def\Stanford{11}
\def\Illinois{12}
\def\Irvine{13}
\def\FII{14}

\def\MPIfR{15}

\shorttitle{\cii\ Emission in SPT0346-52}
\shortauthors{Litke et al.}

\begin{document}

\title{Spatially Resolved \cii\ Emission in SPT0346-52: A Hyper-Starburst Galaxy Merger at $\MakeLowercase{z} \sim 5.7$}

\author{Katrina~C.~Litke$^{\Arizona}$, 
Daniel~P.~Marrone$^{\Arizona}$,
Justin~S.~Spilker$^{\Texas}$, 
Manuel~Aravena$^{\Diego}$, 
Matthieu~B\'ethermin$^{\Marseille}$,
Scott~Chapman$^{\Dalhousie}$,
Chian-Chou~Chen$^{\ESOGarching}$, 
Carlos~de~Breuck$^{\ESOGarching}$,
Chenxing~Dong$^{\UFlorida}$, 
Anthony~Gonzalez$^{\UFlorida}$, 
Thomas~R.~Greve$^{\DAWN,\UCL}$,
Christopher~C.~Hayward$^{\Flatiron}$, 
Yashar~Hezaveh$^{\Stanford}$,
Sreevani~Jarugula$^{\Illinois}$, 
Jingzhe~Ma$^{\Irvine}$, 
Warren~Morningstar$^{\Stanford}$,
Desika~Narayanan$^{\UFlorida,\FII,\DAWN}$, 
Kedar~Phadke$^{\Illinois}$, 
Cassie~Reuter$^{\Illinois}$, 
Joaquin~Vieira$^{\Illinois}$, 
Axel~Wei{\ss}$^{\MPIfR}$
}

\altaffiltext{\Arizona}{Steward Observatory, University of Arizona, 933 North Cherry Avenue, Tucson, AZ 85721, USA; \href{mailto:kclitke@email.arizona.edu}{kclitke@email.arizona.edu}}
\altaffiltext{\Texas}{Department of Astronomy, University of Texas at Austin, 2515 Speedway Stop C1400, Austin, TX 78712, USA}
\altaffiltext{\Diego}{N\'ucleo de Astronom\'ia, Facultad de Ingenier\'ia, Universidad Diego Portales, Av. Ej\'ercito 441, Santiago, Chile}
\altaffiltext{\Marseille}{Aix Marseille Univ., Centre National de la Recherche Scientifique, Laboratoire d’Astrophysique de Marseille, Marseille, France}
\altaffiltext{\Dalhousie}{Dalhousie University, Halifax, Nova Scotia, Canada}
\altaffiltext{\ESOGarching}{European Southern Observatory, Karl Schwarzschild Stra\ss e 2, 85748 Garching, Germany}
\altaffiltext{\UFlorida}{Department of Astronomy, University of Florida, Gainesville, FL 32611, USA}
\altaffiltext{\DAWN}{Cosmic Dawn Center (DAWN), DTU-Space, Technical University of
Denmark, DK-2800 Kgs. Lyngby, Denmark; Niels Bohr Institute,
University of Copenhagen, Juliane Maries vej 30, DK-2100
Copenhagen, Denmark}
\altaffiltext{\UCL}{Department of Physics and Astronomy, University College London, Gower Street, London WC1E 6BT, UK}
\altaffiltext{\Flatiron}{Center for Computational Astrophysics, Flatiron Institute, 162 Fifth Avenue, New York, NY 10010, USA}
\altaffiltext{\Stanford}{Kavli Institute for Particle Astrophysics and Cosmology, Stanford University, Stanford, CA 94305, USA}
\altaffiltext{\Illinois}{Department of Astronomy and Department of Physics, University of Illinois, 1002 West Green St., Urbana, IL 61801}
\altaffiltext{\Irvine}{Department of Physics and Astronomy, University of California, Irvine, CA 92697, USA}
\altaffiltext{\FII}{University of Florida Informatics Institute, 432 Newell Drive, CISE Bldg E251, Gainesville, FL 32611}
\altaffiltext{\MPIfR}{Max-Planck-Institut f\"{u}r Radioastronomie, Auf dem H\"{u}gel 69 D-53121 Bonn, Germany}

\begin{abstract}

SPT0346-52 is one of the most most luminous and intensely star-forming galaxies in the universe, with $\rm L_{\rm FIR} > 10^{13}\ \rm{\lsol}$ and $\Sigma_{\rm SFR} \approx 4200\ \rm{\msol\ yr^{-1}\ kpc^{-2}}$.
In this paper, we present $\sim 0\farcs15$ ALMA observations of the \cii158$\rm{\mu m}$ emission line in this $z = 5.7$ dusty star-forming galaxy.
We use a pixellated lensing reconstruction code to spatially and kinematically resolve the source-plane \cii\ and rest-frame $158\rm{\mu m}$ dust continuum structure  at $\sim 700\ \rm{pc}$ ($\sim 0\farcs12$) resolution.
We discuss the \cii\ deficit with a pixellated study of the $\rm{L_{\cii}/L_{FIR}}$ ratio in the source plane.
We find that individual pixels within the galaxy follow the same trend found using unresolved observations of other galaxies, indicating that the deficit arises on scales $\lesssim 700\ \rm{pc}$.
The lensing reconstruction reveals two spatially and kinematically separated components ($\sim 1\ \rm{kpc}$ and $\sim 500\ \rm{km\ s^{-1}}$ apart) connected by a bridge of gas.
Both components are found to be globally unstable, with Toomre Q instability parameters $\ll 1$ everywhere. 
We argue that SPT0346-52 is undergoing a major merger, which is likely driving the intense and compact star formation.

\end{abstract}


\section{Introduction}

Dusty star-forming galaxies (DSFGs) are among the most infrared-luminous \citep[$\rm{L_{FIR}} > 10^{12}\ \rm{L_{\odot}}$, where $\rm{L_{FIR}}$ is the luminosity integrated from $42.5-122.5\rm{\mu m}$,][]{helou1988} and intensely star-forming ($\Sigma_{\rm SFR} \sim 1000\ \rm{\msol\ yr^{-1}\ kpc^{-2}}$) galaxies in the universe \citep[][]{greve2012,casey2014,ma2015}.
The origin of DSFGs is heavily debated
\citep{sanders1988,engel2010,narayanan2010,hayward2012,hayward2013,chen2015,oteo2016}.
It has been theorized that strong starbursts like DSFGs will eventually form the massive elliptical galaxies seen in the centers of galaxy clusters at $z<1.5$ \citep{thomas2005,thomas2010,kodama2007,kriek2008,zirm2008,gabor2012,hartley2013,toft2014}.

Recent surveys with the 2500 deg$^2$ South Pole Telescope \citep[SPT;][]{vieira2010,carlstrom2011, mocanu2013} have greatly expanded the number of known, bright, strongly lensed DSFGs up to $z\sim 7$ \citep{strandet2017,marrone2018}.
One of the most extreme DSFGs discovered by SPT, with the highest $\rm{L_{FIR}}$ and $\Sigma_{\rm SFR}$ in the SPT sample, is SPT-S J034640-5204.9 (hereafter SPT0346-52).
SPT0346-52 has been studied at radio, infrared, optical, and X-ray wavelengths \citep{vieira2013, weiss2013, hezaveh2013, gullberg2015,ma2015, spilker2015,ma2016, strandet2016,aravena2016,spilker2016}.

SPT0346-52 is a gravitationally lensed galaxy at $z=5.6559$ \citep{weiss2013}, with lensing magnification $\mu = 5.6\pm 0.1$ \citep{spilker2016}.
It has an apparent $\rm{L_{FIR}} = 1.23\times 10^{14}\ \rm{\lsol}$ \citep{spilker2015}
and specific star formation rate $\rm{sSFR} > 15.7\ \rm{Gyr^{-1}}$ \citep{ma2015}.
This galaxy's star formation rate density, $\Sigma_{\rm SFR}$, is $4200\ \rm{\msol\ yr^{-1}\ kpc^{-2}}$, one of the highest of any known galaxy \citep{hezaveh2013,spilker2015,ma2015,ma2016}. 

\cite{hezaveh2013} and \citet{spilker2016} performed gravitational lensing reconstructions of the $860\rm{\mu m}$ continuum emission in SPT0346-52.
This work was continued in \cite{spilker2015}, which 
reconstructed the CO(2-1) line in $200\ \rm{km\ s^{-1}}$ channels.
This lensing reconstruction showed that gas with velocities blueward of $-100\ \rm{km\ s^{-1}}$ was spatially offset from the rest of the emission, but it was unable to distinguish between a merging system of galaxies or a rotation-dominated system due to insufficient spatial resolution ($\gtrsim 0\farcs 5$).

\cite{ma2016} explored the origin  of the high luminosity surface density and star formation rate and found that the infrared luminosity is star-formation dominated, with negligible contributions from a central active galactic nucleus (AGN).

In this paper, we use ALMA Band 7 observations of \cii158$\rm{\mu m}$ (hereafter \cii), a fine-structure line of singly ionized carbon, combined with an interferometric lensing reconstruction tool developed by \cite{hezaveh2016}, to study the structure of SPT0346-52.
\cii158$\rm{\mu m}$ is an ideal tracer of the gas in the interstellar medium (ISM) of galaxies.  At $11.26\ \rm{eV}$, neutral carbon has a lower ionization potential than neutral hydrogen, so \cii\ can be found in many different phases of the ISM and trace regions inaccessible to observations of ionized hydrogen emission.
\cii\ is the dominant cooling line in far-UV heated gas \citep{hollenbach1991}, making it an ideal line with which to study the structure of SPT0346-52.
By studying the structure of this galaxy in \cii\, we can begin to understand what drives the intense star formation rates observed.

The ratio of the \cii\ line luminosity to the far infrared (FIR) continuum luminosity has been observed to decrease as the FIR luminosity increases \citep[e.g., ][]{malhotra1997,luhman1998,luhman2003,diazsantos2013,gullberg2015},
forming the so-called ``\cii\ deficit''
Several different mechanisms to produce the observed \cii\ deficit with respect to $\rm{L_{FIR}}$ have been proposed.
These include charged dust grains in high UV radiation fields, self absorption of \cii\ or optically thick \cii, saturated \cii\ emission in very high density photodissociation regions (PDRs), dust-bounded photoionization regions, and star formation rates driven by gas surface densities \citep{malhotra1997,luhman1998,luhman2003,munoz2016,narayanan2017}.
Pinpointing the origin of this deficit has been difficult, especially since the deficit is not always observed in DSFGs \citep[e.g.,][]{wagg2010,debreuck2014}.
Spatially resolved studies of the \cii\ deficit have recently become possible at high redshift \citep[e.g.,][]{rawle2014,oteo2016} and should be able to provide a more comprehensive look at the gas conditions associated with the deficit.
The analysis in this paper allows for a spatially resolved study of the \cii\ deficit in SPT0346-52.

The \cii\ observations from ALMA are described in Section \ref{sec:obs}.
Section \ref{sec:lens} describes the lensing reconstruction code used and the source-plane reconstruction of SPT0346-52.
The \cii\ deficit and a kinematic analysis of the results are described in Section \ref{sec:analysis} and further discussed in Section \ref{sec:disc}.
A summary and conclusions are provided in Section \ref{sec:conc}.
Throughout this paper, we adopt the cosmology from \cite{planck}, with $\Omega_m = 0.309$, $\Omega_{\Lambda} = 0.691$, and $H_0 = 67.7\ \rm{km\ s^{-1}}$.
At $z=5.6559$, $1'' = 6.035\ \rm{kpc}$ \citep{wright2006}.

\section{ALMA Observations}
\label{sec:obs}

ALMA Band 7 observations of SPT0346-52 were carried out on 2014 September 02 and 2015 June 28 (project ID: 2013.1.01231, PI: Marrone).  SPT0346-52 was observed twice at different resolutions, with $\sim 5$ minutes on source in both observations.    The 2014 September 02 observation used 34 antennae with baselines up to 1.1km.
For the 2015 June 28 observation, 41 antennae were used with baselines up to 1.6km, yielding higher resolution.  The reference frequency (first local oscillator frequency) was 291.53GHz.
J0334-4008 was used as both the flux calibrator and the phase calibrator on both days.
More information about the observations is available in Table \ref{table:obs}.

\begin{figure}
\begin{center}
\includegraphics[width=0.5\textwidth]{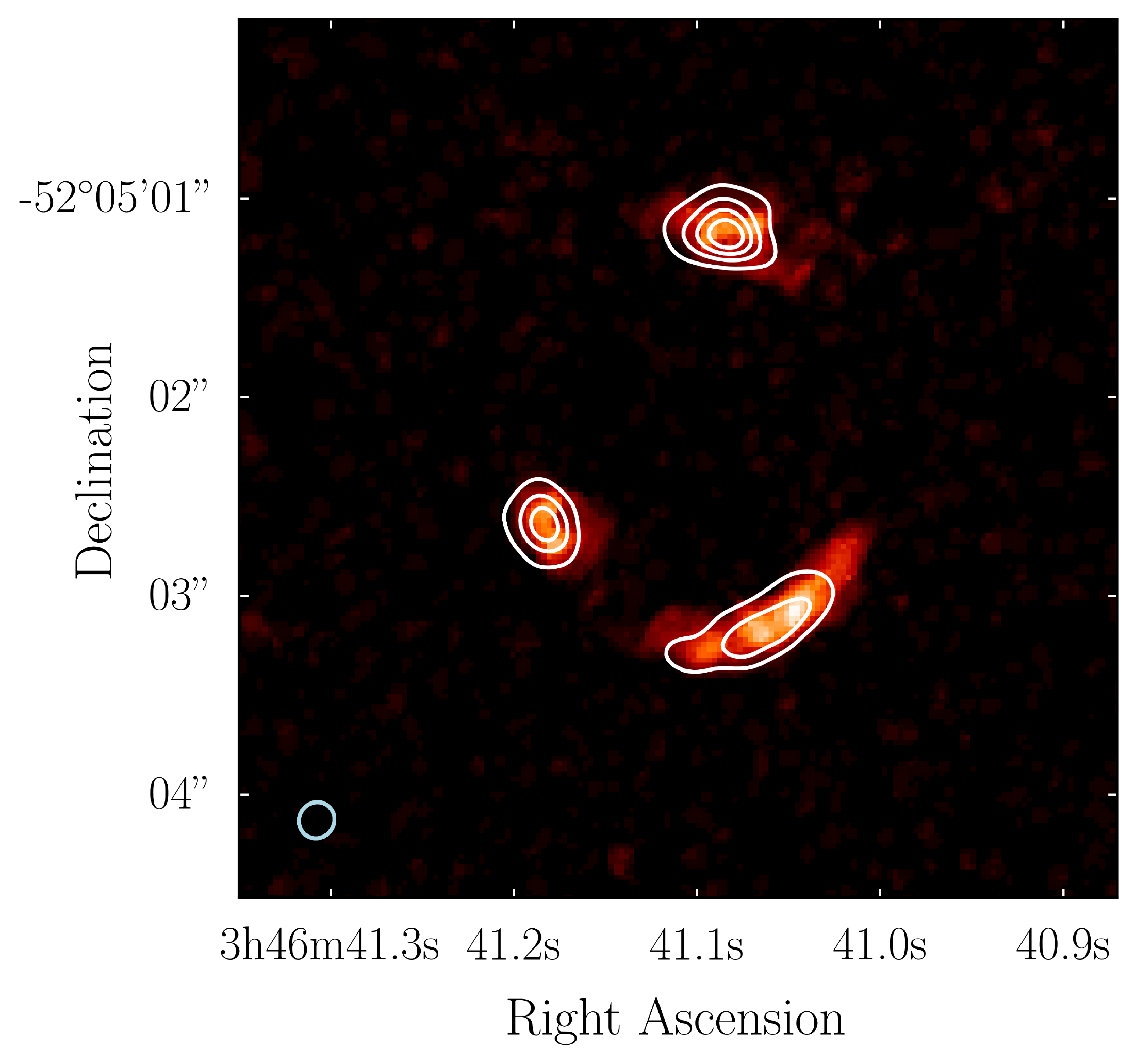}
\caption{High-resolution ALMA observation of SPT0346-52.  The image shows the velocity-integrated \cii\ line.  The continuum is shown overlaid as white contours. The synthesized beam ($0\farcs17\times 0\farcs19$) is illustrated as the bue ellipse in the lower left corner.\label{fig:cont}}
\end{center}
\end{figure}

The data were processed using the Common Astronomy Software Applications package \citep[\textsc{CASA};][]{casa} pipeline version 4.2.2.  Some additional flagging was carried out before processing the data with the pipeline.  Images were made using the \textsc{clean} algorithm within \textsc{CASA}, with Briggs weighting (robust=0.5).  The continuum was subtracted from the line cube using \textsc{uvcontsub} (fitorder=1).  The \cii\ data were binned to $50\ \rm{km\ s^{-1}}$ channels.
The observed $158\rm{\mu m}$ dust emission and integrated \cii\ emission are shown in Figure~\ref{fig:cont}.

\begin{deluxetable*}{lccccc}
\centering
\tablecaption{ALMA Band 7 Observations of SPT0346-52}
\tablenum{1}
\label{table:obs}
\tablehead{\colhead{Date} & \colhead{\# of Ant.} & \colhead{Resolution} & \colhead{PWV$^\mathrm a$} & \colhead{$\mathrm t_{\mathrm{int}}^\mathrm b$} & \colhead{Noise Level$^\mathrm c$}}
\startdata
 &  & (arcsec) &(mm) & (min) & (mJy/beam)  \\
 \hline
2014-Sep-02 & $34$ & $0.26\times 0.23$ & $0.944$ & $5.3$ & $0.44$ \\
2015-Jun-28 & $41$ & $0.19\times 0.17$ & $1.315$ & $5.2$ & $0.54$ 
\enddata
\tablenotetext{a}{Precipitable water vapor at zenith}
\tablenotetext{b}{On-source integration time}
\tablenotetext{c}{Root-mean-square noise level in continuum image}
\end{deluxetable*}

\begin{figure}
\begin{center}
\includegraphics[width=0.5\textwidth]{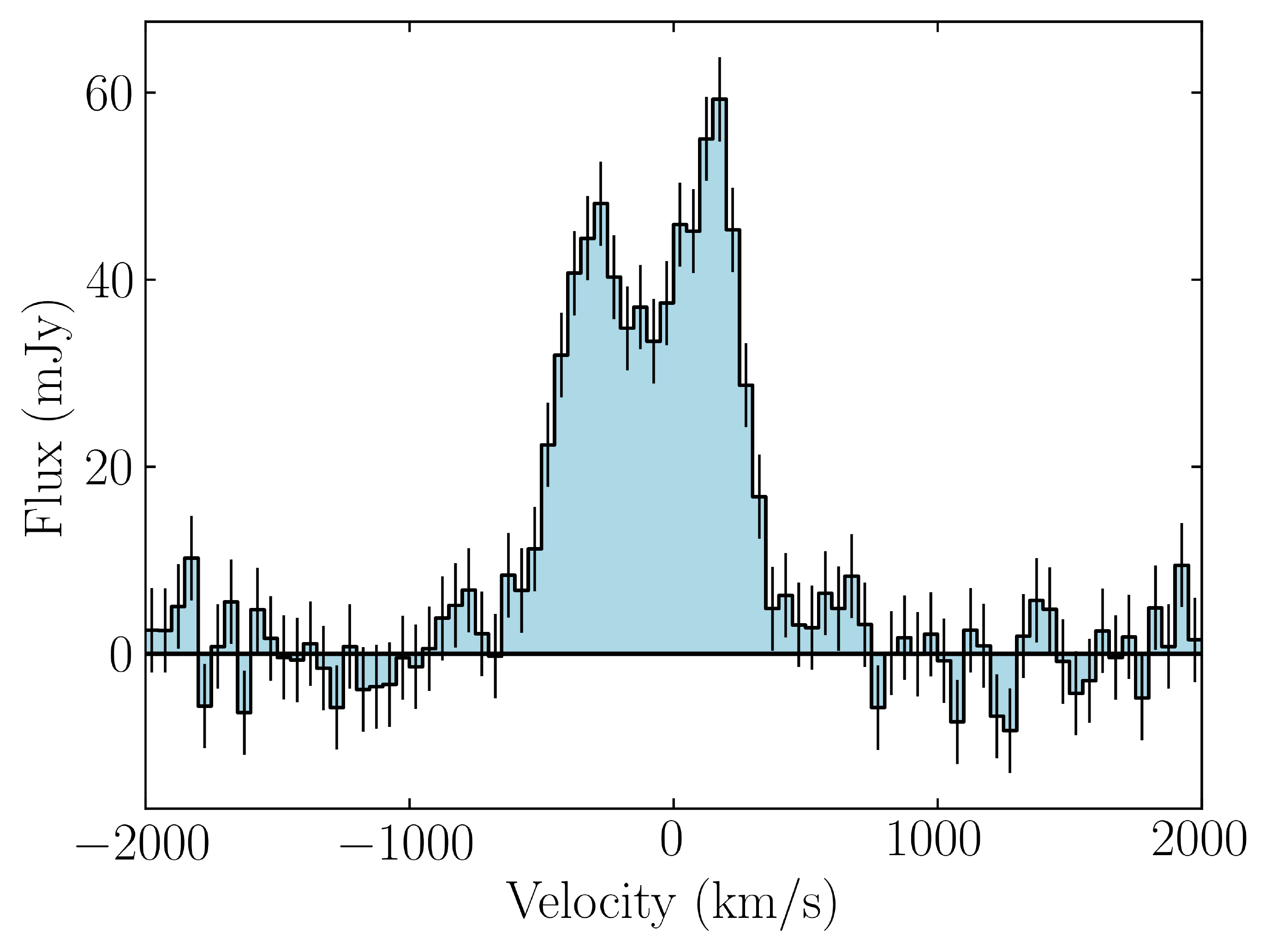}
\caption{Spectrum of observed \cii\ emission in SPT0346-52.  The spectrum was obtained using the observed visibilities and the model continuum visibilities as described in Appendix \ref{app:spec}.\label{fig:spec}}
\end{center}
\end{figure}

The observed \cii\ spectrum is shown in Figure~\ref{fig:spec}.
It shows a profile with two peaks, one red-shifted and one blue-shifted relative to the \cii\ rest frequency.
This spectrum is obtained using the observed complex visibilities rather than \textsc{clean}ed images of the \cii\ line.
The method used to create the spectrum is described further in Appendix \ref{app:spec}.

\section{Lensing Reconstruction}
\label{sec:lens}

Gravitational lensing is a useful phenomenon for observing faint emission. Lensing conserves surface brightness of lensed background sources but it increases their apparent sizes, resulting in greater observed flux. However, strong lensing produces multiple distorted images of background sources and studying the intrinsic properties of lensed objects requires correcting for the lensing distortion.

\subsection{Pixellated Lensing Reconstruction}
\label{sec:ripples}

To determine the structure of SPT0346-52, we use the pixellated lensing reconstruction code \textsc{ripples}.
This code is described in detail in \cite{hezaveh2016}, with the general framework of using pixellated sources described in \cite{warren2003} and \cite{suyu2006}. 
\textsc{ripples} models the interferometric observations of lensed sources.
It models the mass distribution in the lensing galaxy and the background source emission while accounting for observational effects such as those due to the primary beam.

Using a pixellated source reconstruction is advantageous as it does not assume a specific source structure (i.e., the source is not constrained to follow, for example, a Gaussian or S\'ersic surface brightness profile).
Instead, it has the flexibility to model more complex source structures, especially when high-resolution data are available, because of the large parameter space of the source pixels and the less constraining priors.
Using an inherently interferometric code such as \textsc{ripples} also allows us to use all of the data available from an observing session with an interferometer like ALMA. 

The model visibilities can be written as a linear matrix equation,
\begin{equation}
\label{eq:lens}
V = F(BLS).
\end{equation}
Light from the background source, $S$, is first lensed by the foreground galaxy.
Pixels in the image (lensed) plane are mapped back to the source (de-lensed) plane for a given set of lens parameters.
The lensing operation is a matrix represented by $L$ in Equation \ref{eq:lens}, and depends on the mass distribution of the lensing galaxy.
The lensed emission is then modified by the primary beam of the telescope (represented by the matrix $B$).
Finally, we take a Fourier transform of the sky emission, $F$ to obtain the complex visibilities of interferometric observations, $V$.
The model visibilities are compared to the observed visibilities via a $\chi^2$ goodness-of-fit test.
A Markov Chain Monte Carlo (MCMC) method is used to solve for the lens galaxy mass distribution parameters.

In addition to the lens parameters, there is a regularization term, $\lambda$.
The regularization term acts to smooth the source and minimize large gradients between adjacent pixels in the source plane.
This prevents over-fitting of the data, or fitting to the noise in the source plane image.
$\lambda$ is determined by
\begin{equation}
N_s-\lambda Tr([FL+\hat{C_s}]^{-1}\hat{C_s}^{-1})-\lambda S^T\hat{C_s}S = 0,
\end{equation}
where $N_s$ is the number of source pixels and $C_s$ is the source covariance matrix.
$\lambda$ scales an arbitrarily normalized source covariance matrix, $\hat{C}_s$.
It is determined for a fixed lens model, rather than being simultaneously fit for with the lens parameters.
We fit for $\lambda$, then run the MCMC with \textsc{ripples}.  These two steps are repeated until the chains have converged around a most-likely set of parameters.

After modeling the mass distribution of the lensing galaxy, we obtain a pixellated map of the source-plane emission, a model image, and model complex visibilities.

\subsection{Reconstruction of SPT0346-52}

\begin{figure*}
\begin{center}
\includegraphics[width=0.85\textwidth]{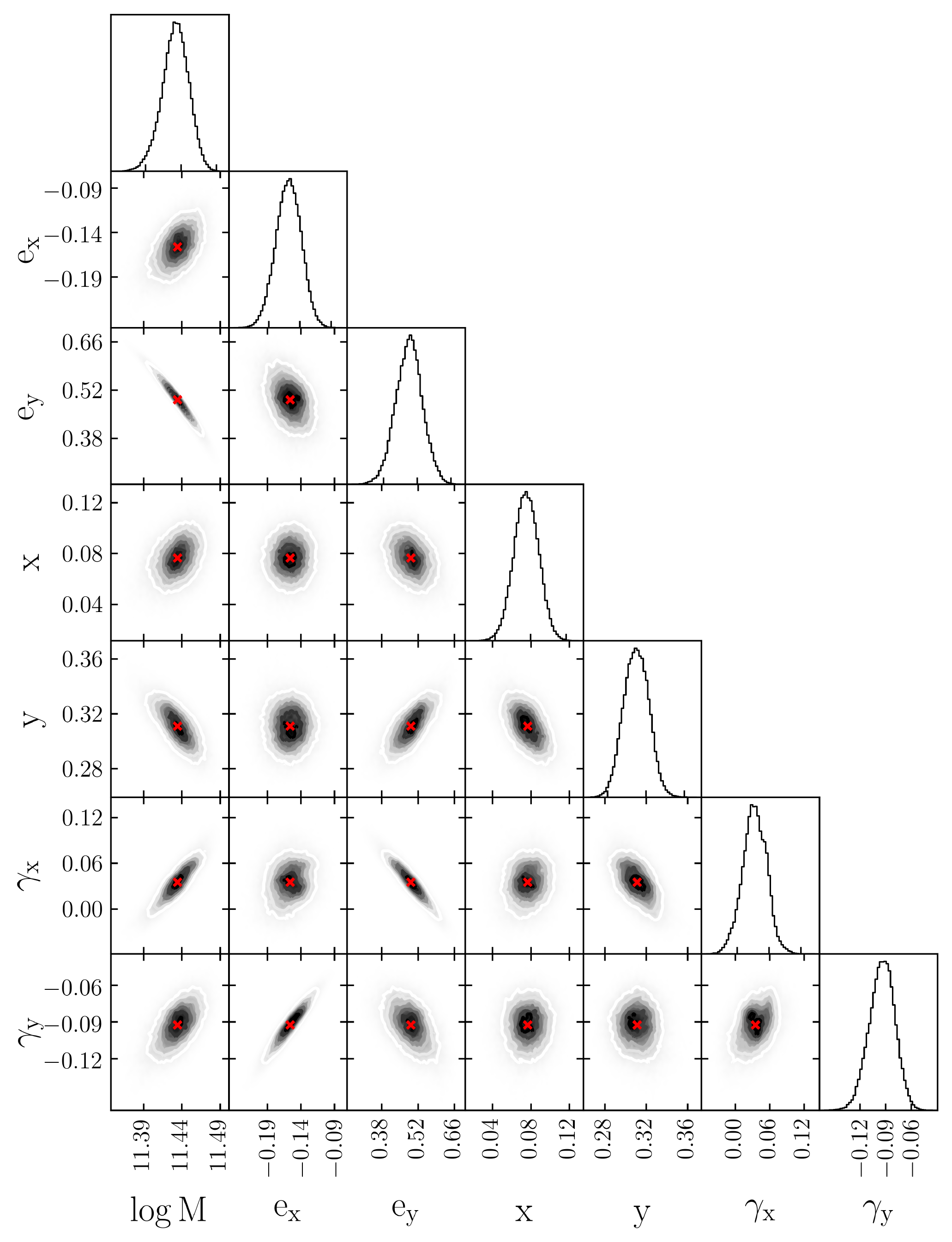}
\caption{Triangle plot with the model lens parameters from reconstructing SPT0346-52.
$M$ is measured in $\msol$ and is the mass enclosed within $10\ \rm{kpc}$.  $e_x$ and $e_y$ are the $x$- and $y$- components of lens galaxy's ellipticity.
$x$ and $y$ are the offset of the lens center in arcseconds.
$\gamma_x$ and $\gamma_y$ are the $x$- and $y$- components of shear.  \label{fig:triangle}}
\end{center}
\end{figure*}

We model the lensing galaxy as a singular isothermal ellipsoid at $z=0.9$ with an external shear component.
The initial parameters are taken from previous lensing reconstructions of SPT0346-52 by \cite{hezaveh2013} and \cite{spilker2015}.
The best-fit model was determined by fitting the $158\ \rm{\mu m}$ continuum data because the continuum has a much higher signal-to-noise ratio than the individual line channels.
Figure~\ref{fig:triangle} shows a probability density plot of the lens parameters with the results of the MCMC.
The determined lens parameters are given in Table \ref{table:mcmc}.

\begin{figure*}
\begin{center}
\includegraphics[width=0.45\textwidth]{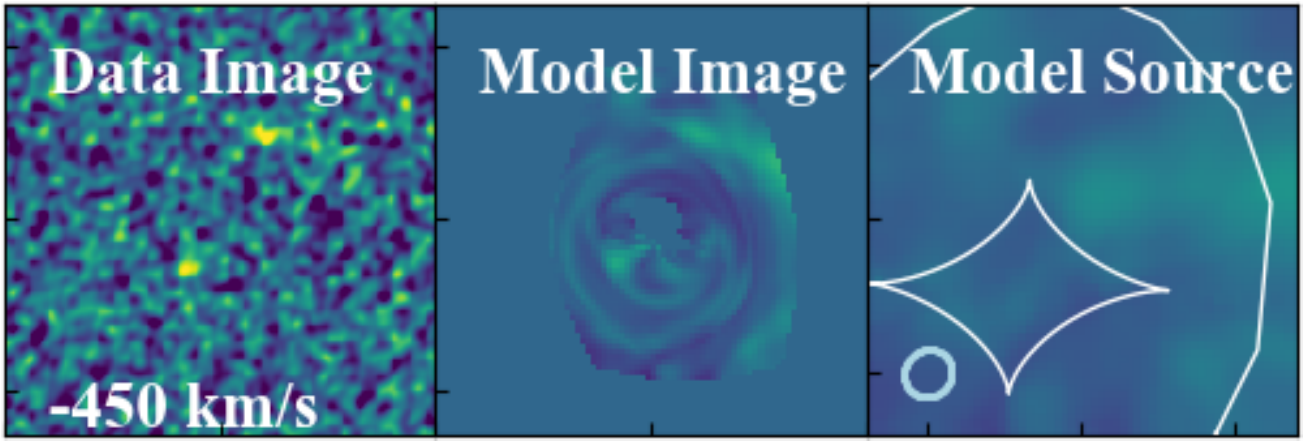}
\includegraphics[width=0.45\textwidth]{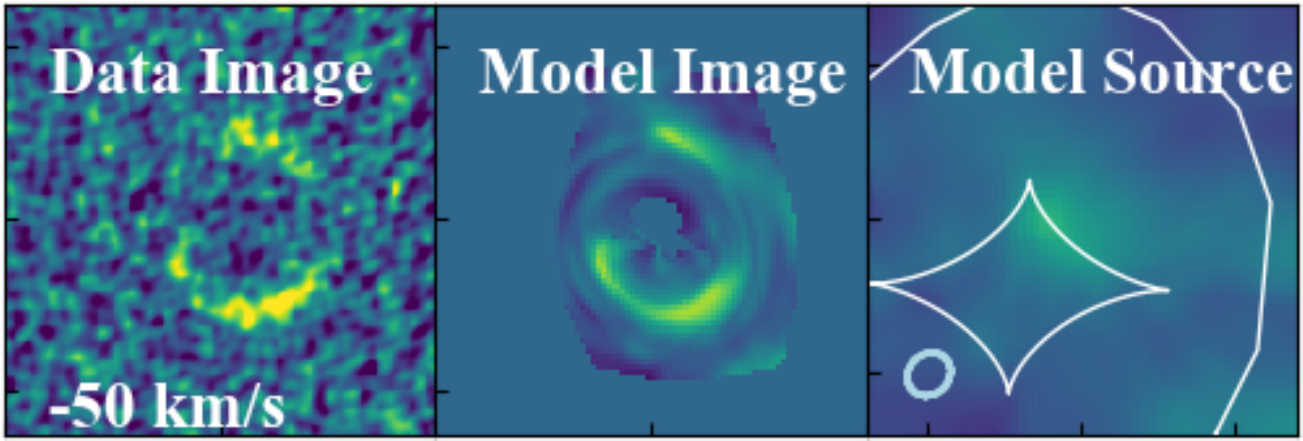}
\includegraphics[width=0.45\textwidth]{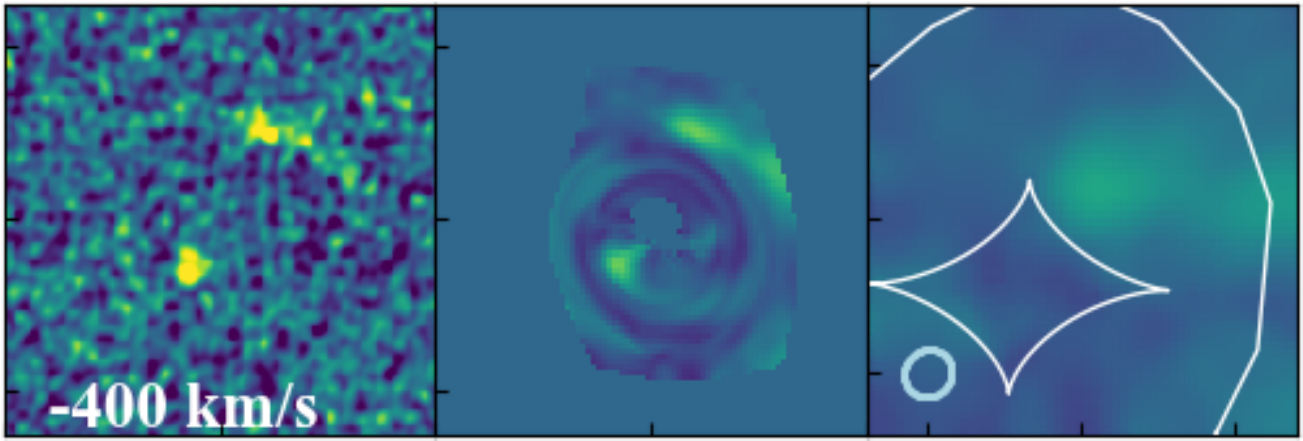}
\includegraphics[width=0.45\textwidth]{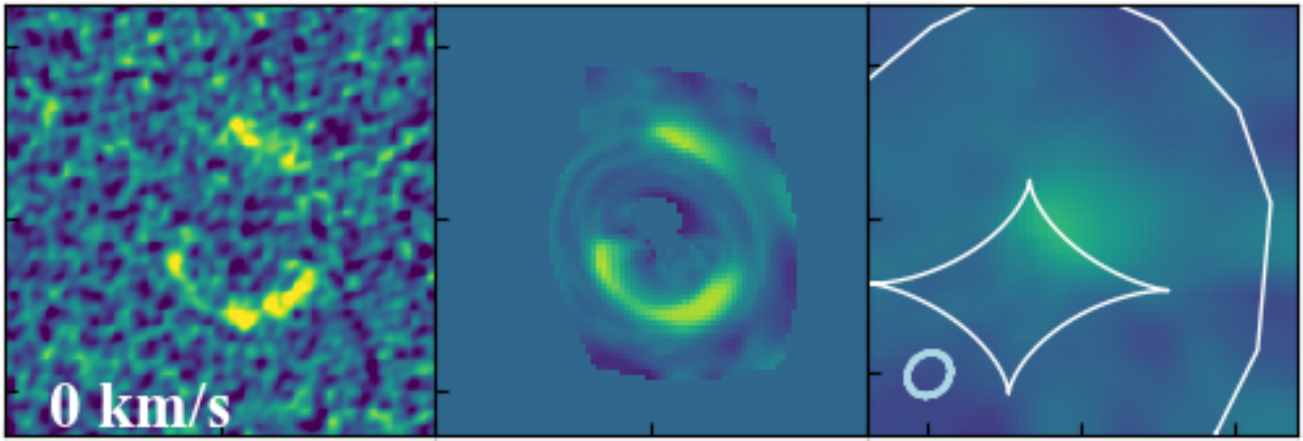}
\includegraphics[width=0.45\textwidth]{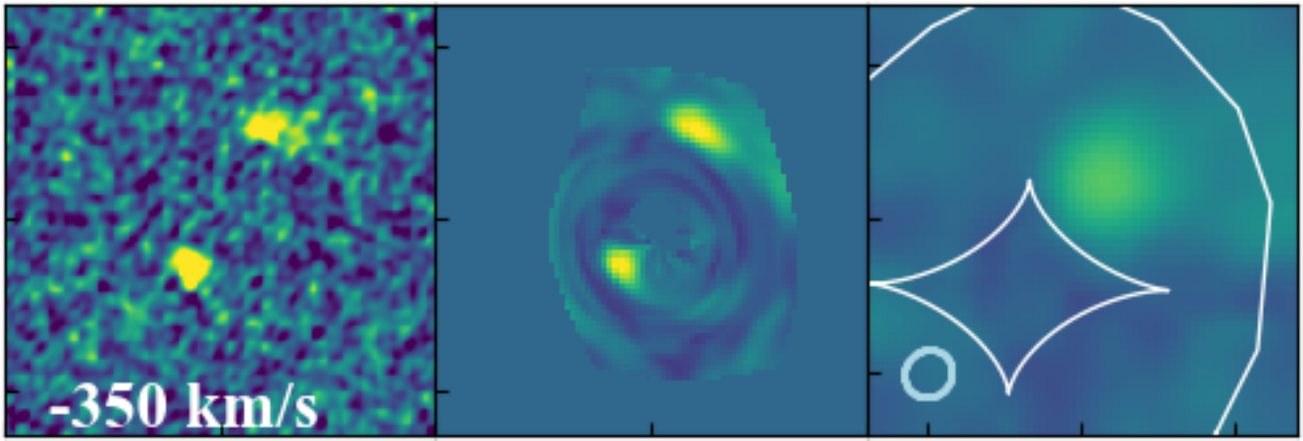}
\includegraphics[width=0.45\textwidth]{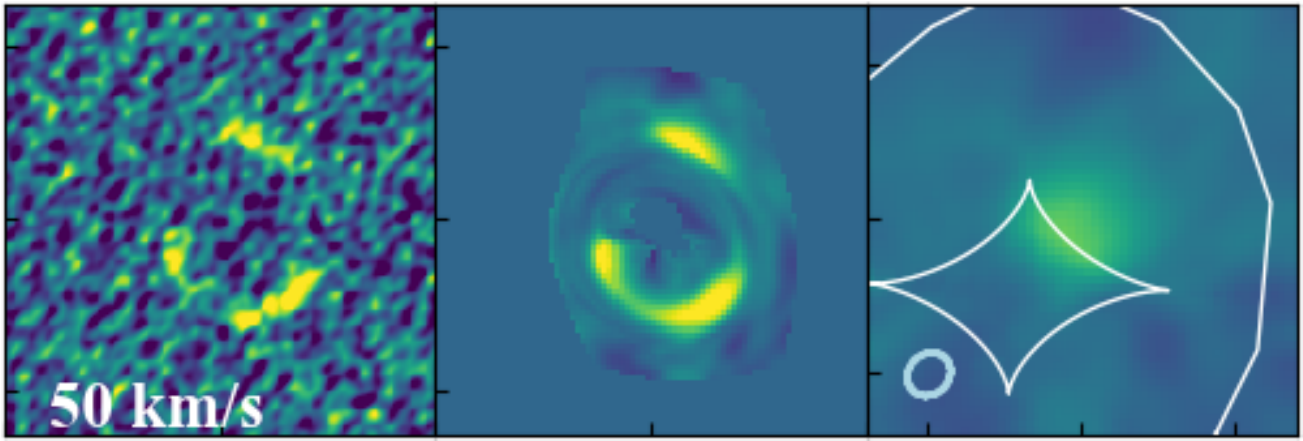}
\includegraphics[width=0.45\textwidth]{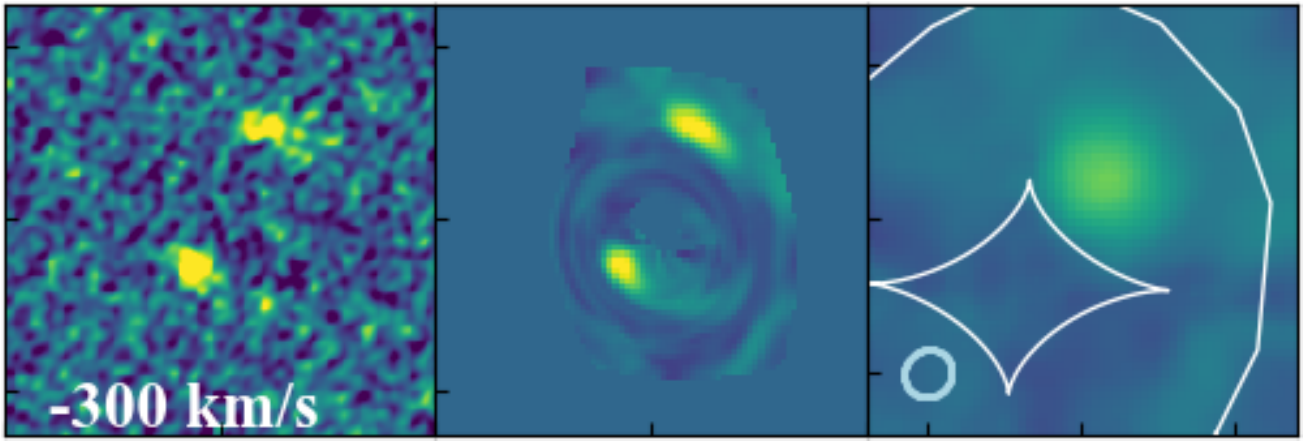}
\includegraphics[width=0.45\textwidth]{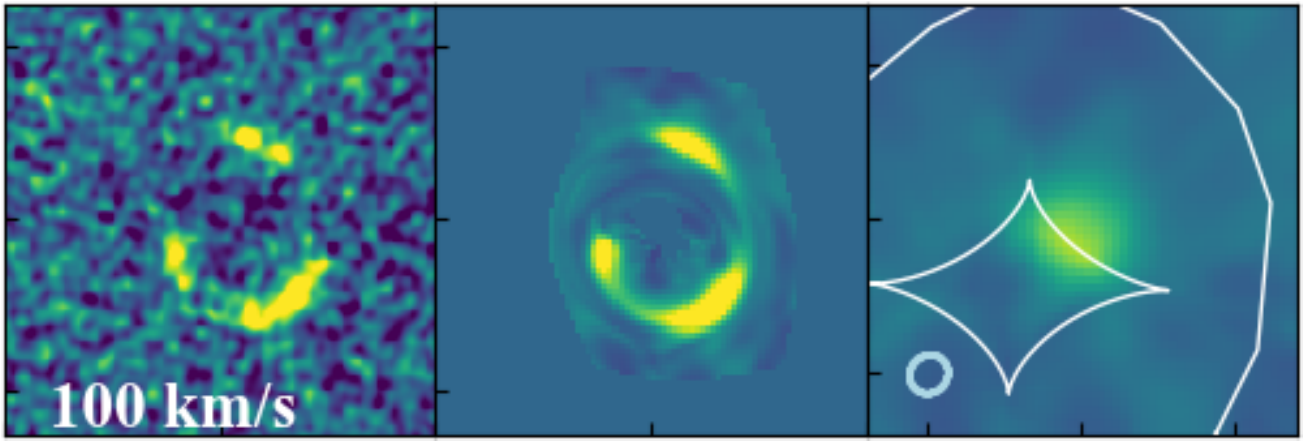}
\includegraphics[width=0.45\textwidth]{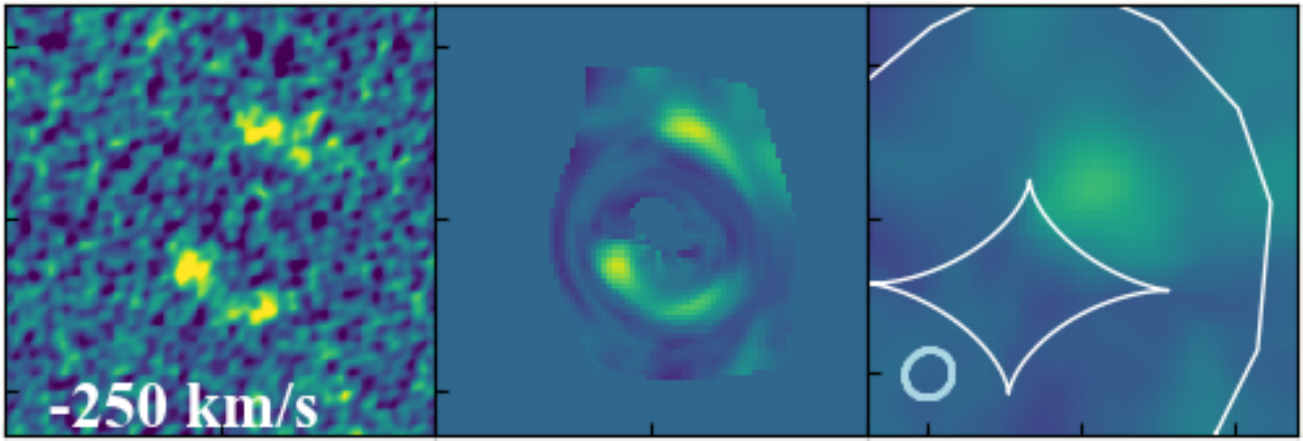}
\includegraphics[width=0.45\textwidth]{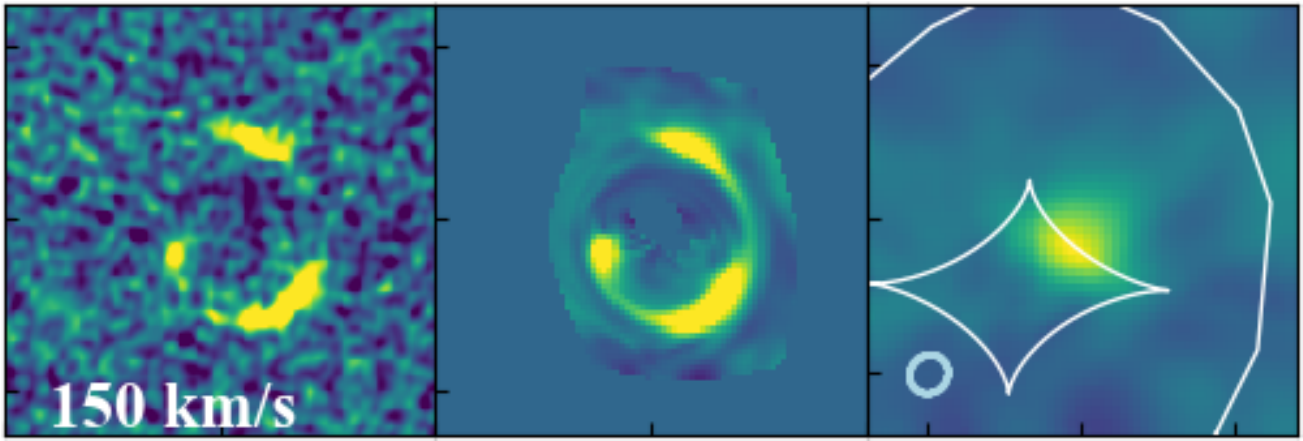}
\includegraphics[width=0.45\textwidth]{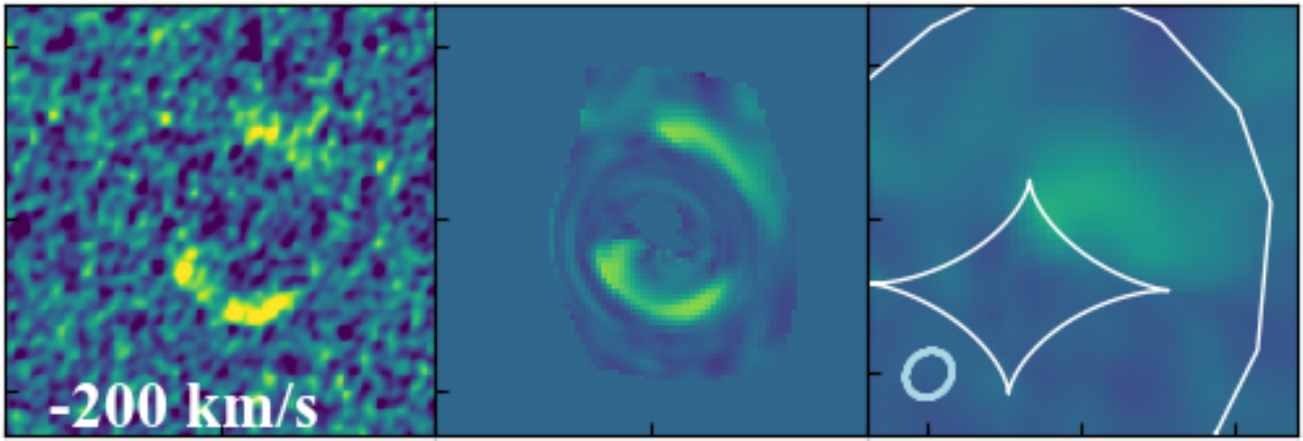}
\includegraphics[width=0.45\textwidth]{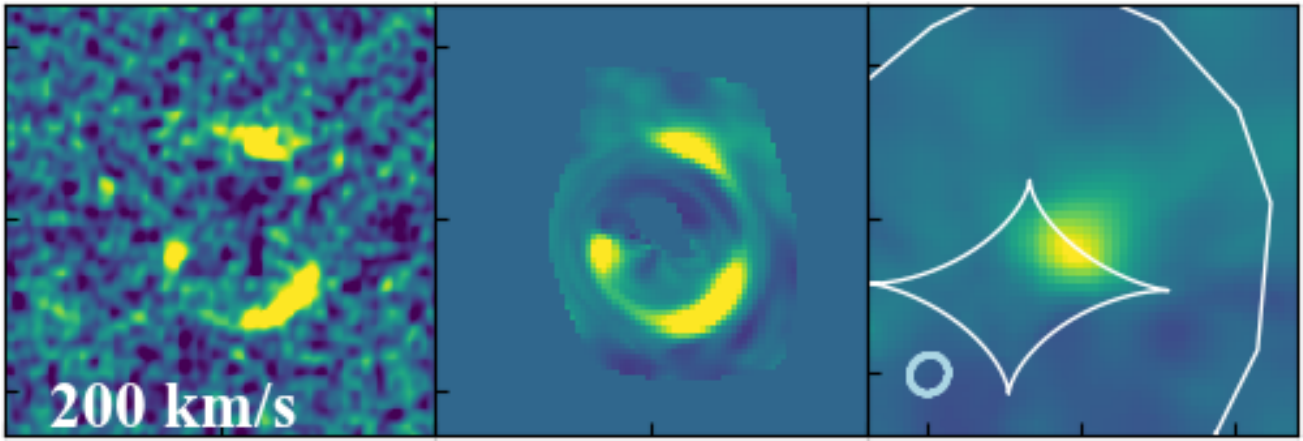}
\includegraphics[width=0.45\textwidth]{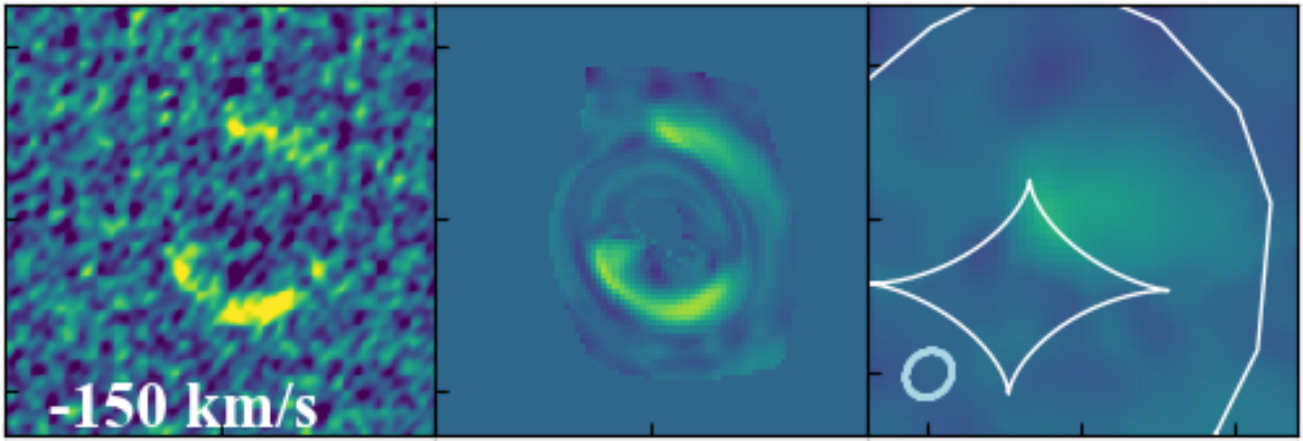}
\includegraphics[width=0.45\textwidth]{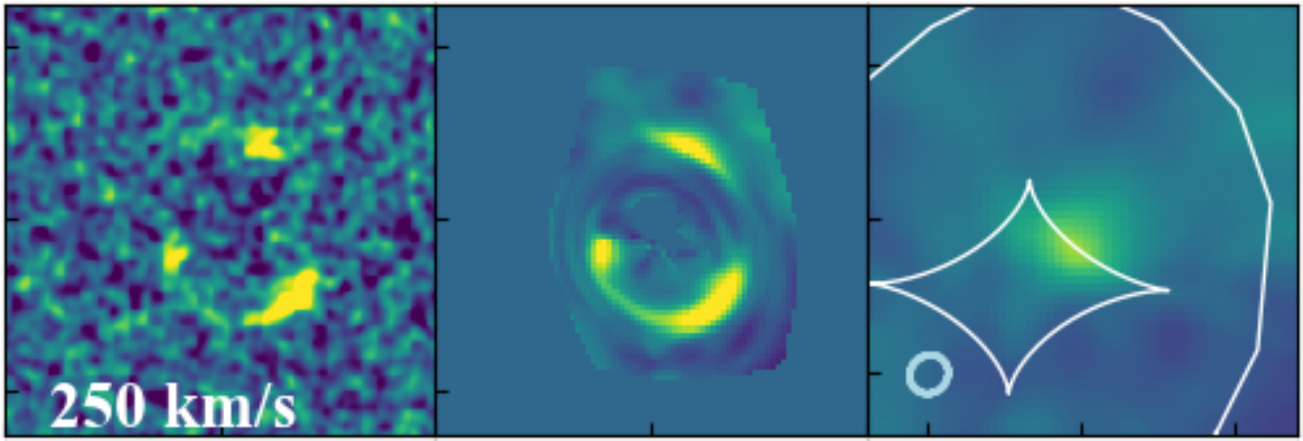}
\includegraphics[width=0.45\textwidth]{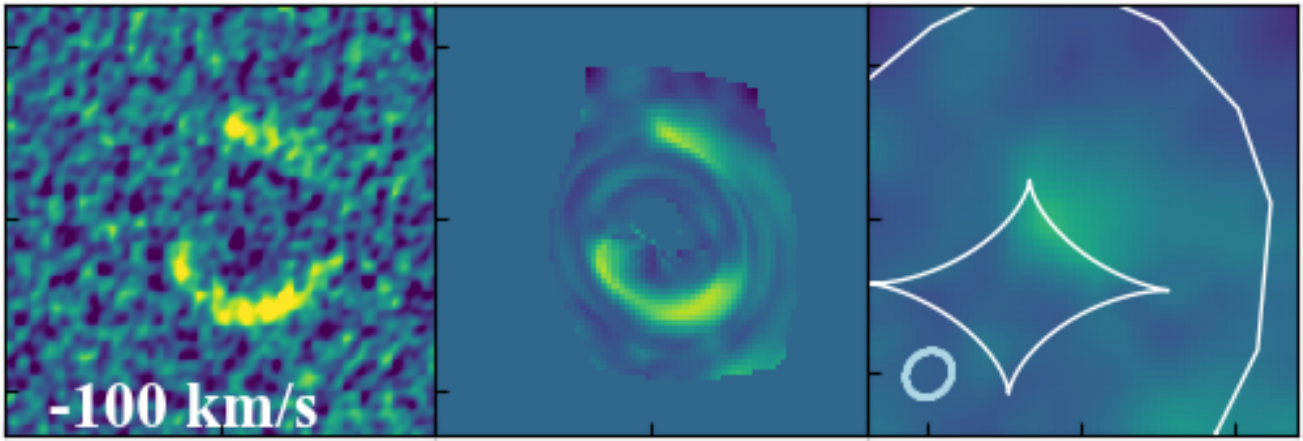}
\includegraphics[width=0.45\textwidth]{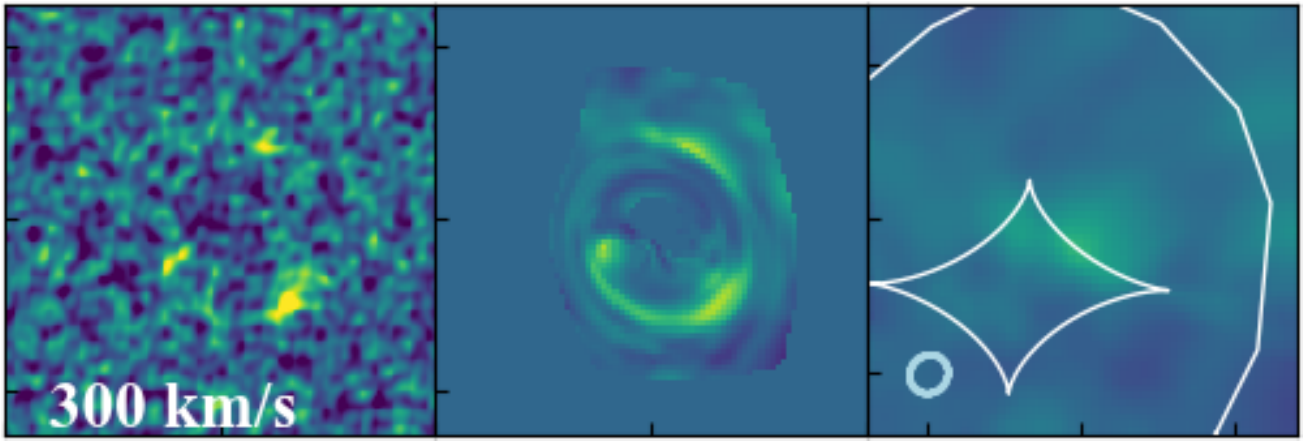}
\caption{Reconstruction of \cii$158\rm{\mu m}$ emission in $50\ \rm{km\ s^{-1}}$ channels.  Left 3-panel column:  blue velocities.  Right 3-panel column:  red velocities.  Within each column:  Left:  observed \cii\ from ALMA.  Center:  model sky emission.  Right:  reconstructed source with lensing caustic. Lensed images are $5''$ on a side.  The source plane images are $8.5\ \rm{kpc}$ a side.  The light blue ellipses in the corners of the source-plane panels show the effective resolution where the emission is brightest in each channel. The data and model sky emission have one color scale, and images of the reconstructed source have another color scale.  These color scales are the same across all channels.
\label{fig:panel}}
\end{center}
\end{figure*}

The best-fit model was then applied to the \cii\ line in each $50\ \rm{km\ s^{-1}}$ channel.
This channel width was chosen to be wide enough to have high signal-to-noise to be able to reconstruct the source in each channel, while being narrow enough to study kinematic features in SPT0346-52.
The source regularization, $\lambda$, for the \cii\ line was fixed for all $50\ \rm{km\ s^{-1}}$ channels.
The original image, model image, and model source are shown for each channel in Figure~\ref{fig:panel}.

\begin{deluxetable}{rc}
\tablecaption{SPT0346-52 Lens Parameters}
\tablenum{2}
\label{table:mcmc}
\tablehead{\colhead{Parameter} & \colhead{Value}} \\
\hline
\startdata
$\rm{\log{Mass[\msol]}}$ & $11.43 \pm 0.02$\\
Ellipticity x-Component, $e_x$ & $-0.16 \pm 0.02$\\
Ellipticity y-Component, $e_y$ & $0.49 \pm 0.06$\\
Ellipticity, $e^{\mathrm a, c}$ & $0.52 \pm 0.06$ \\
Position Angle, $\phi_e$ (E of N)$^{\mathrm b, c}$ & $72 \pm 6^{\circ}$ \\
Lens x Position, $x$ & $0\farcs076 \pm 0\farcs014$\\
Lens y Position, $y$ & $0\farcs31 \pm 0\farcs01$\\
Shear x-Component, $\gamma_x$ & $0.035 \pm 0.022$\\
Shear y-Component, $\gamma_y$ & $-0.093 \pm 0.015$\\
Shear Amplitude, $\gamma^{\mathrm a, c}$ & $0.10 \pm 0.02$ \\
Shear Position Angle, $\phi_{\gamma}$ (E of N)$^{\mathrm b, c}$ & $111 \pm 12^{\circ}$ 
\enddata
\tablenotetext{a}{$\alpha = \sqrt{\alpha_x^2+\alpha_y^2}$, where $\alpha = e$ or $\alpha = \gamma$}
\tablenotetext{b}{$\phi_{\alpha} = \arctan{(-\alpha_y/\alpha_x)}$, where $\alpha= e$ or $\alpha = \gamma$}
\tablenotetext{c}{Derived from best-fit parameters}
\end{deluxetable}

\cite{hezaveh2013} reconstructed the $860\ \rm{\mu m}$ continuum of SPT0346-52 only using short baselines, assuming a symmetric Gaussian source profile.
\cite{spilker2015} reconstructed the CO(2-1) line emission using the code \textsc{visilens} \citep{spilker2016}.
They used four $200\ \rm{km\ s^{-1}}$ channels and assumed a symmetric Gaussian source-plane structure for each channel.
Channels blueward of $-100\ \rm{km\ s^{-1}}$ were spatially offset from redder emission, with the same orientation and velocity ranges obtained with the reconstruction of the \cii\ line.
The $-400\ \rm{km\ s^{-1}}$ and $+200\ \rm{km\ s^{-1}}$ channels from the parametric reconstruction of CO(2-1) in \cite{spilker2016} show disks with similar size and orientation as the two components in the \cii\ pixellated reconstruction from this work (see Figure~\ref{fig:panel}).

\begin{figure}
\begin{center}
\includegraphics[width=0.4\textwidth]{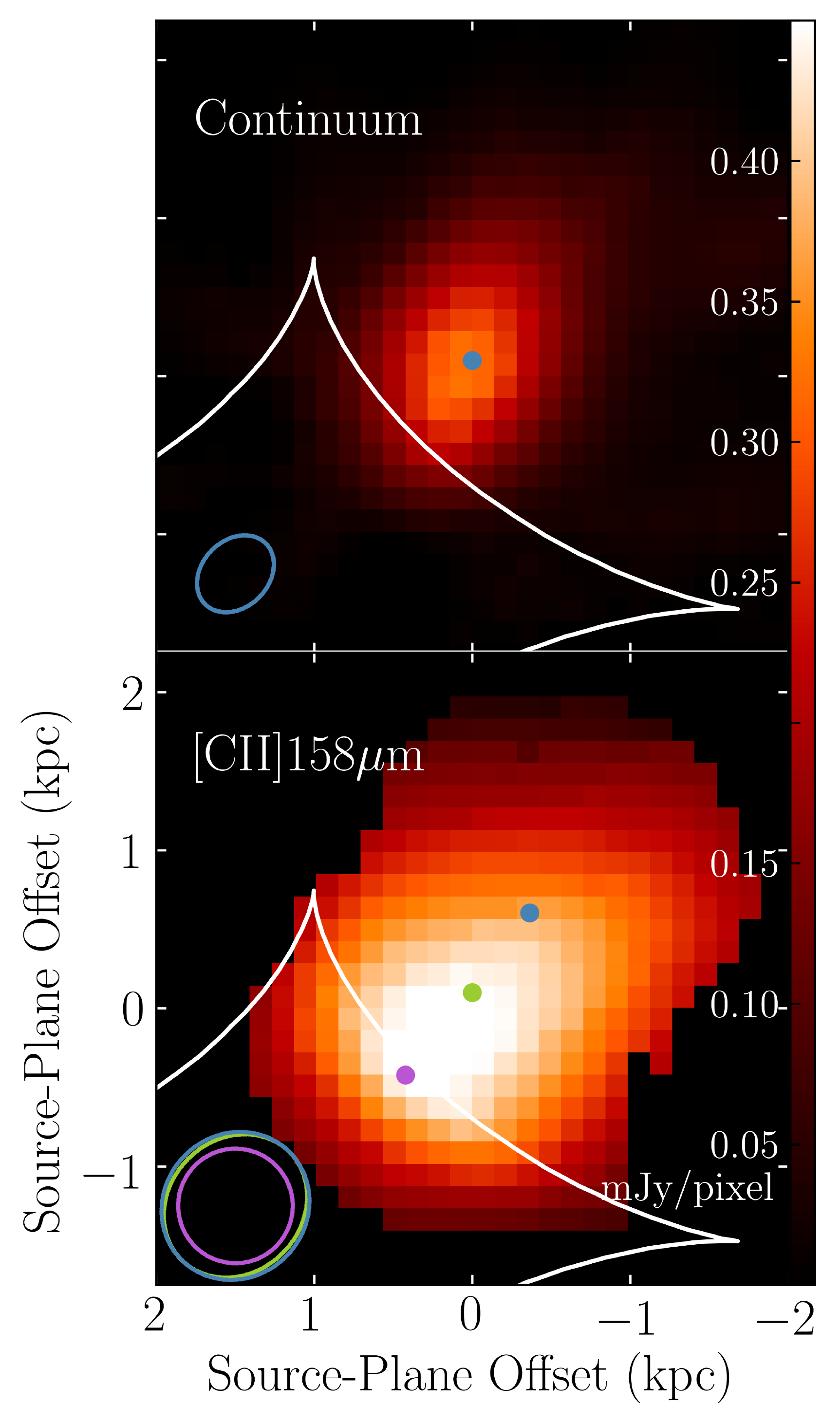}
\caption{Result of pixellated lensing reconstruction of SPT0346-52.  Top: map of continuum emission at $158\rm{\mu m}$.  Bottom:  map of integrated \cii\ emission.  
The resolution varies across the source depending on the location relative to the caustics (white lines). The blue, green, and purple ellipses in the corner indicate the 1$\sigma$ size of the 2D Gaussian fit to determine the effective resolution in the reconstruction at the locations of the dots in the maps of the same color. 
To see more of how the resolution varies across the source plane, see Figure \ref{fig:panel}.
Colorbar units are mJy/pixel.
\label{fig:source}}
\end{center}
\end{figure}

We combined the reconstructed channels from Figure~\ref{fig:panel} into a map of the \cii\ emission (Figure~\ref{fig:source}).  The top image shows the reconstructed continuum emission, while the bottom image shows the velocity-integrated reconstructed \cii\ line.
There are two lobes in the \cii\ emission, with the lower left component much brighter than the upper right component.
The continuum emission, arising mostly from dust, is more regular and is roughly elliptical.
It is located near the center of the \cii\ emission, between the two components and is less extended than the \cii\ emission.
The dust continuum emission has been found to be more compact than the \cii\ emission in other high-z galaxies.
\cite{gullberg2018} measured \cii-emitting regions that are $1.6\times$ more extended than the regions with dust continuum emission in four $z\sim 4.5$ DSFGs.
\cite{wang2013} found dust continuum in $1.2-2.3\ \rm{kpc}$ regions and \cii\ emission in $1.7-3.5\ \rm{kpc}$ regions in $z\sim6$ quasars with vigorous star formation in the central region of the quasar host galaxies.
\cite{oteo2016} found similar sizes in the dust continuum and \cii\ emission regions of SGP38326.
An offset between the brightest \cii\ emission and the center of the dust emission, as is seen in the source-plane reconstruction of SPT0346-52, was also observed in the Seyfert 2 galaxy NGC1068 \citep{herreracamus2018a}.
We also find the dust continuum emission to be smoother than the \cii\ emission.
Smooth dust continuum emission with clumpy \cii\ emission was also seen by \cite{oteo2016} in SGP38326, a pair of interacting dusty starbursts.

\begin{figure}
\begin{center}
\includegraphics[width=0.45\textwidth]{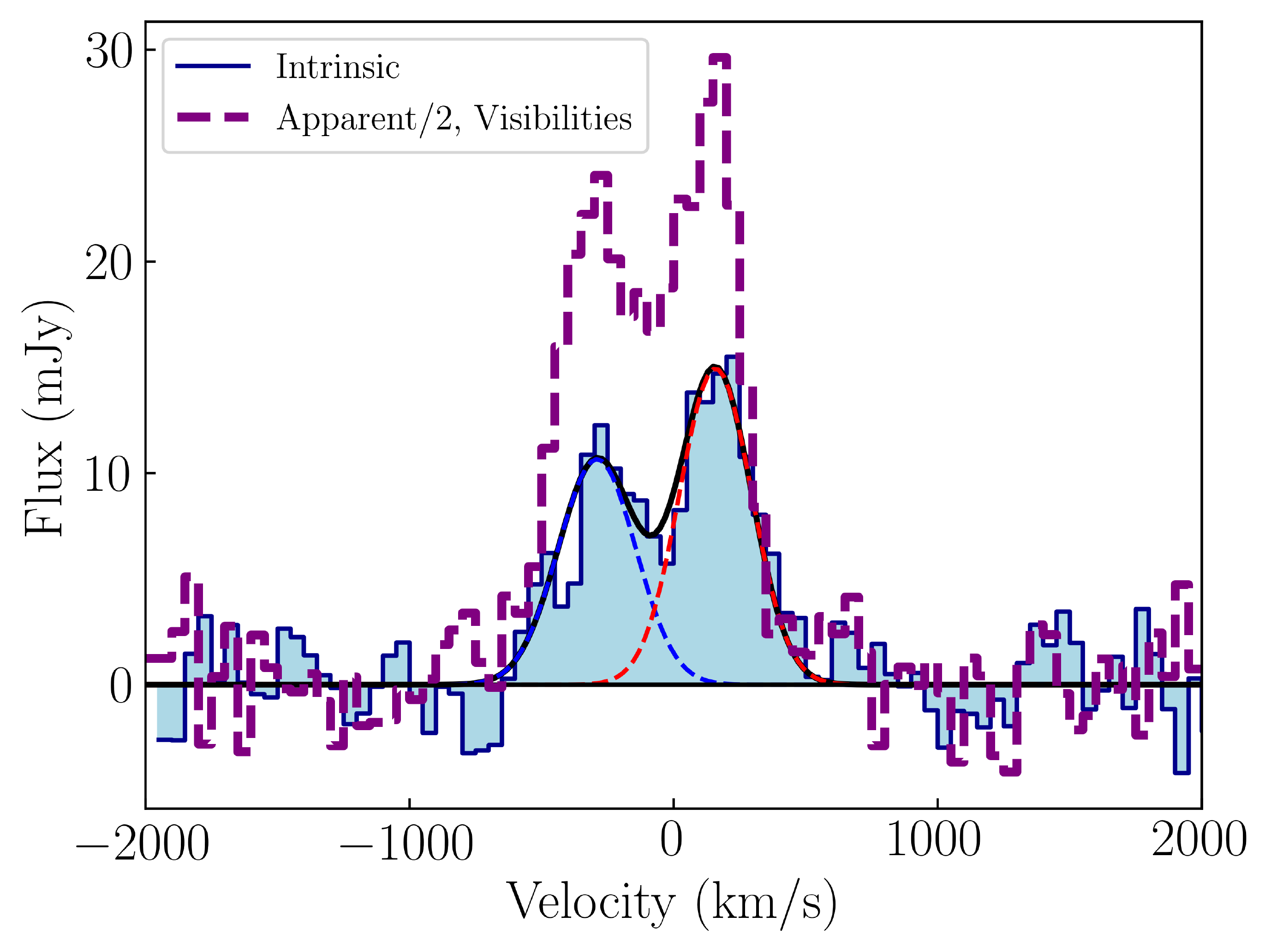}
\caption{Spectrum of reconstructed \cii\ emission in SPT0346-52.  The solid black line shows the two component Gaussian fit to the spectrum, with the individual components shown in red and blue dashed lines.  The two components are centered at $-287\ \rm{km\ s^{-1}}$ and $+158\ \rm{km\ s^{-1}}$ relative to $z=5.6559$.  
The purple dashed line shows the spectrum obtained from the (lensed) visibilities in Figure~\ref{fig:spec}.
\label{fig:srcspec}}
\end{center}
\end{figure}

Figure~\ref{fig:srcspec} shows a spectrum of the reconstructed \cii\ emission.  The spectrum was obtained by summing the flux from the pixels in the source plane reconstruction, while excluding pixels near the edges of the source plane.
There are two clear peaks in the spectrum with similar maximum fluxes.
A two-component Gaussian was fit to the spectrum; the results are overlaid on Figure~\ref{fig:srcspec}.  The blue component is centered at $-287\pm 22\ \rm{km\ s^{-1}}$ and has a FWHM of $337\pm 22\ \rm{km\ s^{-1}}$.
The red component is centered at $+158 \pm 22\ \rm{km\ s^{-1}}$ with a FWHM of $319 \pm 15\ \rm{km\ s^{-1}}$.
Similar velocity structure to that seen in the source-plane spectrum was also observed by \cite{spilker2015}, \cite{aravena2016}, \citet[][submitted]{dong2018}, and Apostolovski et al. (in prep).

\subsection{Source Plane Resolution}
\label{sec:res}

In order to determine the resolution in the source plane, we created a set of mock visibilities for a point source in the source plane that was lensed by the best-fit lens model.
\textsc{ripples} was then applied to the mock visibilities to reconstruct the source and a 2D Gaussian was fit to the reconstructed source.
This Gaussian is the effective resolution.
This process was repeated with the location of the point source varying throughout the source plane to understand the variation in resolution across the source. We find that in regions away from the central diamond caustic the effective resolution is $0\farcs 13 \times 0\farcs 15$, while
closer to the diamond caustic the effective resolution decreases to $0\farcs 12 \times 0\farcs 12$.
Example effective resolution ellipses are shown in blue in Figure~\ref{fig:source}.

\section{Analysis}
\label{sec:analysis}

\subsection{\cii\ Deficit}
\label{sec:cii}

\cii\ is usually the brightest coolant line of the ISM.
While it can be emitted in a variety of ISM conditions, it is primarily produced in warm, diffuse gas at the edges of photodissociation regions (PDRs) being heated by an external FUV radiation field, such as a star-forming region or AGN \citep{hollenbach1991,malhotra1997,luhman1998,pineda2010}.
\cite{pavesi2018} calculated that $\sim 85\%$ of \cii\ emission in a $z=5.7$ DSFG comes from PDRs.
One of the more interesting aspects of the \cii line is the so-called ``\cii\ deficit'', in which the $\rm{L_{\cii}/L_{FIR}}$ ratio has been found to decrease at high $\rm{L_{FIR}}$, though this is not always the case.
The deficit is often associated with AGN activity, though not all AGN have a \cii\ deficit \citep{sargsyan2012}.
\cite{farrah2013} also showed that the deficit is stronger in merging systems, with no clear dependence on the presence of an AGN.
The deficit was found to be strongest in AGN with the highest central starlight intensities, rather than those with the highest X-ray luminosities at low redshift \citep{smith2017}.
This is further supported by \cite{lagache2017}, who found that the \cii\ deficit is correlated with the interstellar radiation field in their simulations.
In resolved \cii\ studies of the Orion Nebula in our galaxy and other DSFGs, the \cii\ deficit has been shown to be strongest in regions with higher star formation rates \citep{goicoechea2015,oteo2016}.

Several mechanisms have been suggested as a cause for the \cii\ deficit.
The \cii\ line may be optically thick or self-absorbed by foreground gas.
Enhanced IR emission, from intense star formation or an AGN, can also lead to a deficit \citep{malhotra1997,luhman1998}.
More recently, \cite{narayanan2017} proposed that increased surface densities in clouds and increased star formation rates cause a rise in the fraction of gas that is CO-dominated, rather than \cii-dominated, leading to a \cii\ deficit.
In addition, \cite{diazsantos2017} found a correlation between the UV flux to gas density ratio, $G/n_H$, and $\Sigma_{\rm IR}$.
They found a critical surface density, $\Sigma_{\rm IR}^* \simeq 5\times 10^{10}\ \rm{\lsol\ kpc^{-2}}$, below which $G/n_H$ remains constant.
Above $\Sigma_{\rm IR}^*$, they found that $G/n_H$ increases.
They argued that the relation between $G/n_H$ and $\Sigma_{IR}$ links kpc-scale galaxy properties to those of individual PDRs. 
\cite{herreracamus2018a} also found a critical surface density, $\Sigma_{\rm FIR} \gtrsim 10^{11}\ \rm{\lsol}$, above which the $\rm{L_{\cii}/L_{FIR}}$ ratio decreases, but with increased scatter.
The \cii\ deficit has also been found to correlate directly with $G/n_H$ \citep[e.g.,][]{malhotra1997}.

Using the pixellated lensing reconstruction, we have resolved maps of the source-plane continuum and \cii\ emission.
This allows us to obtain a resolved map of the $\rm{L_{\cii}/L_{FIR}}$ ratio and probe the \cii\ deficit to smaller scales than has previously been possible at high redshift.

In order to study the \cii\ deficit, we assume that the $158\rm{\mu m}$ continuum flux density, $\rm{F_{cont}}$, traces the $\rm{L_{FIR}}$ ratio.
The measured continuum flux in each pixel, $\rm{F_{cont,i}}$, is scaled proportional to the total $\rm{L_{FIR}}$, using  $\rm{L_{FIR}}$ from \cite{gullberg2015} and corrected for lensing using the magnification from \cite{spilker2016} such that
\begin{equation}
\rm{L_{FIR,i}} = \rm{{F_{cont,i}}} \left( \frac{\rm{L_{FIR}}}{\rm{\Sigma F_{cont_i}}} \right).
\end{equation}

\noindent This method assumes a constant flux density-to-luminosity ratio, and thus a constant dust temperature.
We tested the effect of this assumption by determining the total FIR luminosity for a range of dust temperatures measured in DSFGs from the SPT sample ($22 < T_d < 57$). 
$\rm{L_{FIR}}$ was calculated by integrating the spectral energy distribution (SED; modeled by a modified black body, see \citealt{greve2012}) from $42.5-122.5 \mu m$ and scaling the SED to go through the flux of SPT0346-52 at $158\rm{\mu m}$.
Resulting $\rm{L_{FIR}}$ values were within a factor of $\sim 2$ of the luminosity measured by \cite{gullberg2015}, so the variation caused by variable dust temperatures in the galaxy is within a factor of $\sim 2$.

\begin{figure}
\begin{center}
\includegraphics[width=0.4\textwidth]{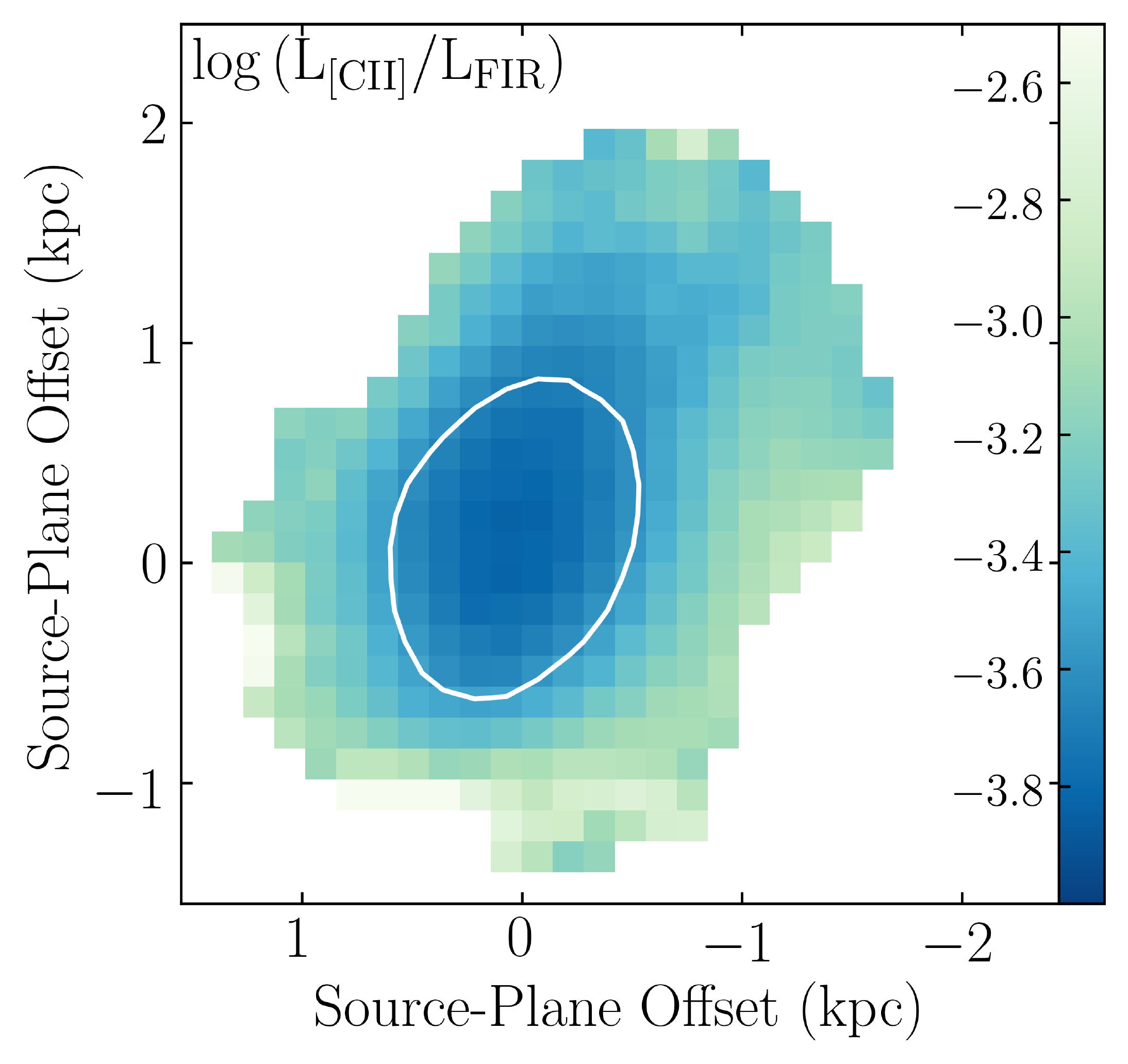}
\caption{\cii to FIR luminosity ratio in SPT0346-52.  The center region, where the continuum emission is strongest, shows relatively uniform values for $\rm{L_{\cii}/L_{FIR}}$ in the center, $\sim 1 \times 10^{-4}$.  The white contour traces the strongest region of continuum emission.  The larger values at the edges of the galaxy are due to the falling continuum emission. \label{fig:ratio}}
\end{center}
\end{figure}

A map of the $\rm{L_{\cii}/L_{FIR}}$ ratio is shown in Figure~\ref{fig:ratio}.
Typical values of the $\rm{L_{\cii}/L_{FIR}}$ ratio in the center of SPT0346-52 are around $\sim 1 \times 10^{-4}$.
This value is consistent with other ultra-luminous infrared galaxies (ULIRGs) and DSFGs that have the \cii\ deficit \citep[e.g.,][]{maiolino2005,iono2006,oteo2016,mazzucchelli2017,decarli2017}.
The higher values of $\rm{L_{\cii}/L_{FIR}}$ at the edges of the galaxy are due to the low amounts of continuum emission in those regions.
\cite{oteo2016} found a similar mapped distribution in a pair of interacting DSFGs and suggested it was due to the different morphology of the \cii\ emission compared to the dust continuum emission.
As with SPT0346-52, the sources studied by \cite{oteo2016} do not show evidence for AGN activity \citep{oteo2016}.
The uniformity of the $\rm{L_{\cii}/L_{FIR}}$ ratio is similar to the merging system observed by \cite{neri2014}.

\begin{figure*}
\begin{minipage}{1.\textwidth}
\begin{center}
\includegraphics[width=0.45\textwidth]{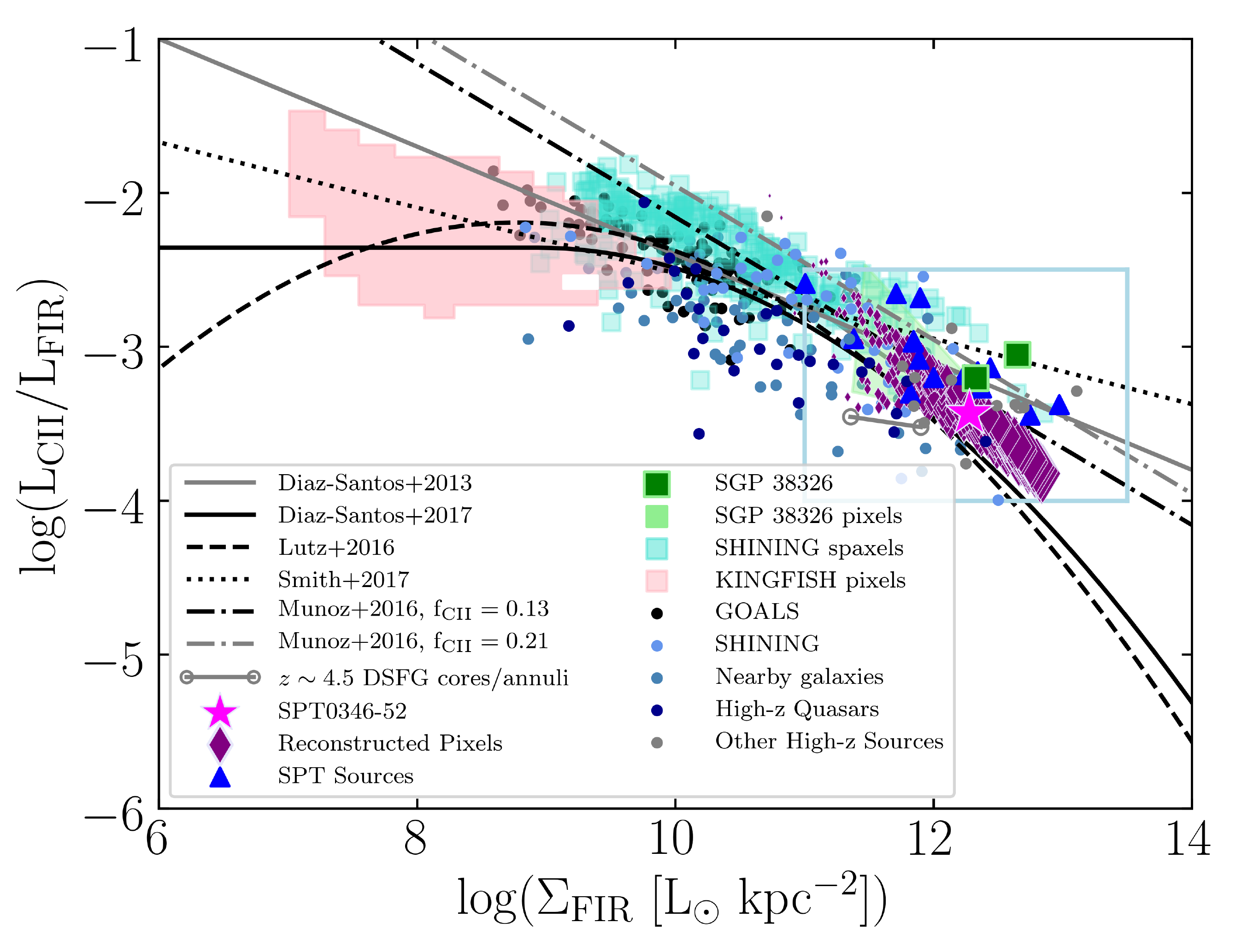}
\includegraphics[width=0.45\textwidth]{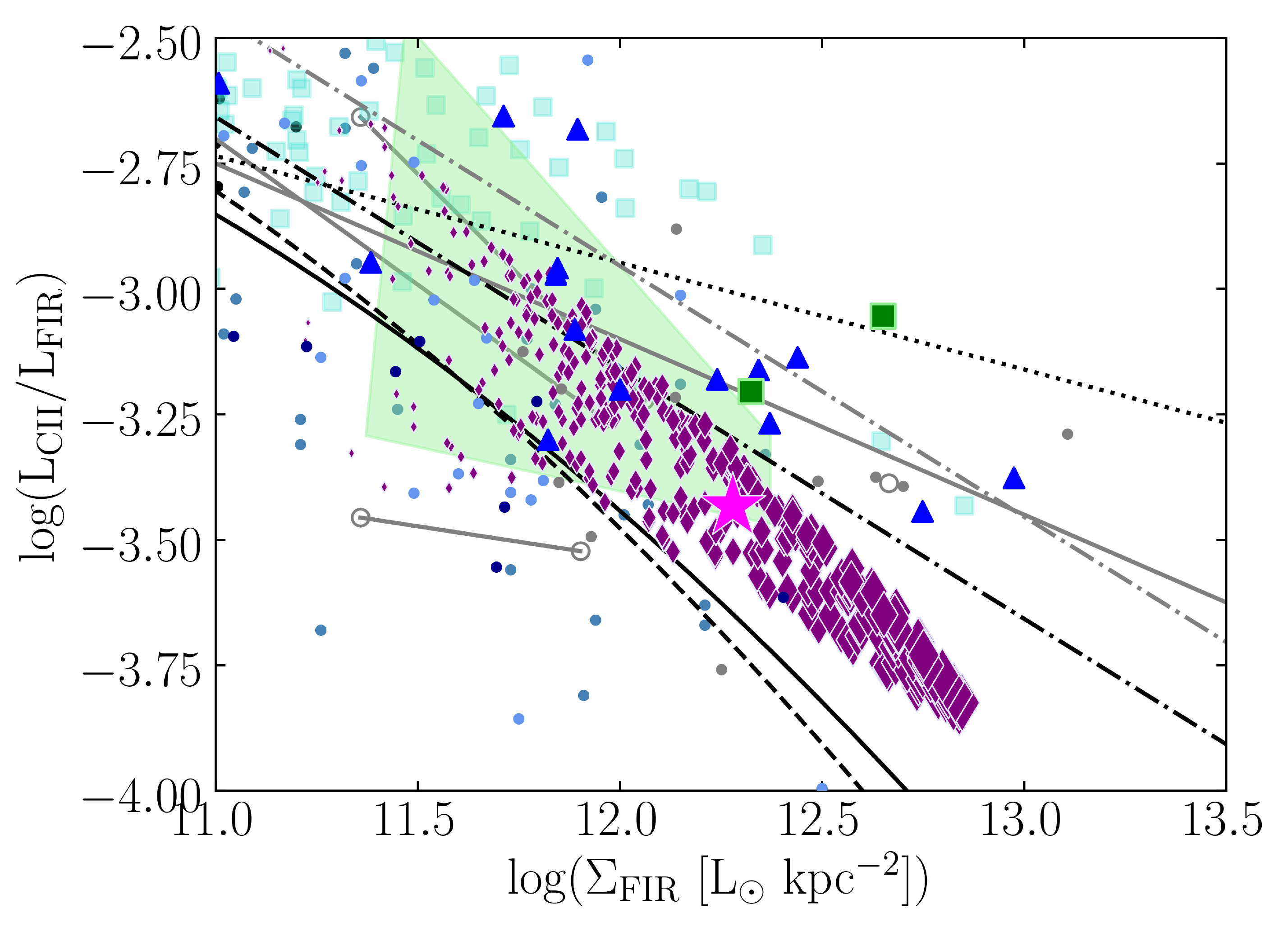}
\caption{$\rm{L_{\cii}/L_{FIR}}$ vs $\Sigma_{\rm FIR}$.  The magenta star shows the galaxy-summed value for SPT0346-52 from this work.  The purple diamonds show the individual pixels from the lensing reconstruction.  The size of the diamonds is weighted by the continuum signal-to-noise.
  The right panel zooms in on the region from the left panel with the pixels from the SPT0346-52 reconstruction (light blue box). 
The green squares are the resolved DSFGs (SGP 38326) from \cite{oteo2016}, while the green shaded region outlines the parameter space occupied by their individual pixels.
The turquoise shaded squares represent spaxels from a spatially resolved survey of FIR lines in the $z\sim 0$ \textsc{shining} sample \citep{herreracamus2018a}.
The pink shaded region represents binned pixels from a spatially resolved study of the \cii\ deficit in the \textsc{kingfish} galaxy sample \citep{smith2017}.
  The blue triangles are SPT DSFGs \citep{gullberg2015}.  
The black points are low-redshift star-forming galaxies and ULIRGs from the \textsc{goals} survey \citep{diazsantos2013}.  
The light blue points represent global values from nearby galaxies in the \textsc{shining} survey \citep{herreracamus2018a}.
The steel blue points are additional nearby galaxies \citep{farrah2013,brauher2008}.
  The dark blue points are $z>6$ quasars from \cite{decarli2018} and \cite{izumi2018}.
The grey points are additional high-redshift objects from the literature \citep{walter2009,carniani2013,riechers2013,wang2013,debreuck2014,neri2014,riechers2014,yun2015,diazsantos2016,pavesi2018,hashimoto2018}.
The grey circles represent the $\rm{L_{\cii}/L_{FIR}}$ ratio in the cores (measured within the continuum spectrum) and annuli (measured between the continuum and \cii\ apertures) for three DSFGs at $z\sim 4.5$ in \cite{gullberg2018}.  The solid grey lines connect core of a given DSFG to the annulus of that DSFG.  The core values have higher $\Sigma_{\rm FIR}$ and lower $\rm{L_{\cii}/L_{FIR}}$ than their annulus counterparts.
  The grey solid line represents the relation found by \cite{diazsantos2013} and extended by \cite{spilker2016}.  
  The black dash-dot line represents the relation from \cite{munoz2016} saturation of the \cii\ line for $\rm{f_{\cii}=0.13}$, the value for typical DSFGs.
The grey dash-dot line represents the relation from \cite{munoz2016} for $\rm{f_{\cii}=0.21}$, the value calculated for SPT0346-52.
  The solid line is the relation found by \cite{diazsantos2017}.
  The dotted line is the relation fit by \cite{smith2017}
  $\Sigma_{\rm SFR}$ values were converted to $\Sigma_{\rm FIR}$ following \cite{murphy2011} and \cite{greve2012}. The dashed line is the fit from \cite{lutz2016}.
  \label{fig:surf}}
\end{center}
\end{minipage}
\end{figure*}

Figure~\ref{fig:surf} shows the $\rm{L_{\cii}/L_{FIR}}$ vs $\Sigma_{\rm FIR}$ relation for SPT0346-52, pixels from the lensing reconstruction, other high-z sources, and ULIRGs from the \textsc{goals} survey \citep{diazsantos2013}.
As noted by \cite{spilker2016}, the $\rm{L_{\cii}/L_{FIR}}$ vs $\Sigma_{\rm FIR}$ relation continues to higher values of $\Sigma_{\rm FIR}$ for high-z sources.
The tight relation continues to hold true at smaller physical scales (the purple diamonds and green shaded region in Figure~\ref{fig:surf} are individual pixels in resolved \cii\ observations from this work and \citealt{oteo2016}), with a similar scatter as previous, galaxy-averaged studies at high redshift.

The spatially resolved \cii\ deficit was recently explored in nearby galaxies.  \cite{smith2017} measured $\rm{L_{\cii}/L_{TIR}}$ at $0.2-1.6$ kpc scales in the \textsc{Kingfish} sample and found that the $\rm{L_{\cii}/L_{TIR}}$ vs $\Sigma_{\rm SFR}$ relation continues at lower values of $\Sigma_{\rm SFR}$.
The relation between $\rm{L_{\cii}/L_{TIR}}$ and $\Sigma_{\rm SFR}$ found by \cite{smith2017} is in good agreement with the star formation rate density and $\rm{L_{\cii}/L_{FIR}}$ ratio in SPT0346-52.
This relation, and the similar $\rm{L_{\cii}/L_{FIR}}$ vs $\Sigma_{\rm FIR}$ trend explored in this work, spans many orders of magnitude.
It holds true both for spatially resolved regions and galaxy-averaged values at high and low redshift.
The \cii\ deficit appears to come from local conditions in the ISM because it continues to hold over smaller physical areas.
\cite{gullberg2018} reached the same conclusion in their resolved study of \cii\ emission in four $z\sim4.5$ DSFGs.

\cite{munoz2016} argue that the \cii\ deficit is the result of thermal saturation of the \cii\ emission line.
The relation of $\rm{L_{\cii}/L_{FIR}}$ vs $\Sigma_{IR}$ from \cite{munoz2016} is plotted in Figure~\ref{fig:surf} for $f_{\rm \cii} \approx 0.13$ as a black dash-dotted line.
$f_{\rm\cii}$ is proportional to $\rm{L_{\cii}/L_{FIR}}$, and $f_{\rm \cii} \approx 0.13$ is the fraction of the total gas in a galaxy traced by \cii\ for a typical DSFG.
However, if we calculate $f_{\rm \cii}$ (Equation 5 from \citealt{munoz2016}) for SPT0346-52 using $\alpha_{\rm CO} = 2.2$, using $L'_{\rm CO(2-1)}$ from \citet{spilker2015} and assuming $L'_{\rm CO(1-0)} = L'_{\rm CO(2-1)}$, we find that $\rm f_{\cii} = 0.21$.
This moves the relation from \cite{munoz2016} above the majority of the pixels in the reconstruction of SPT0346-52 in Figure~\ref{fig:surf} (grey dash-dot line).
It should be noted that the other lines shown in Figure~\ref{fig:surf} are empirical fits to the data.

\cite{herreracamus2018a} looked at the $\rm{L_{\cii}/L_{FIR}}$ ratio in the \textsc{shining} sample of nearby galaxies, with spatially resolved information for 25 of their galaxies.
In \cite{herreracamus2018b}, they use a pair of toy models to explore the origin of the \cii\ deficit, one with the ISM modeled as having OB stars and molecular gas clouds closely related, and the other with OB associations and neutral gas clouds randomly distributed throughout the ISM.
In the former case, the \cii\ intensity only weakly depends on $G_0$ (because the ionization parameter reaches a limit, $U\approx 0.01$) and $n_{\rm H}$ (because the density of the neutral gas exceeds the critical density for collisional excitation of \cii), 
and in the latter case, the \cii\ intensity is nearly independent of $G_0$ (because photoelectric heating efficiency decreases), 
while the FIR intensity is proportional to $G_0$ in both scenarios.
They conclude that the combination of both scenarios best replicates the observed \cii\ deficit, including a critical luminosity surface density of $\Sigma_{\rm FIR} \approx 10^{10}\ \rm{\lsol\ kpc^{-2}}$ above which the $\rm{L_{\cii}/L_{FIR}}$ ratio begins to decline.

\subsection{Kinematic Analysis}
\label{sec:kin}

\begin{figure}
\begin{center}
\includegraphics[width=0.45\textwidth]{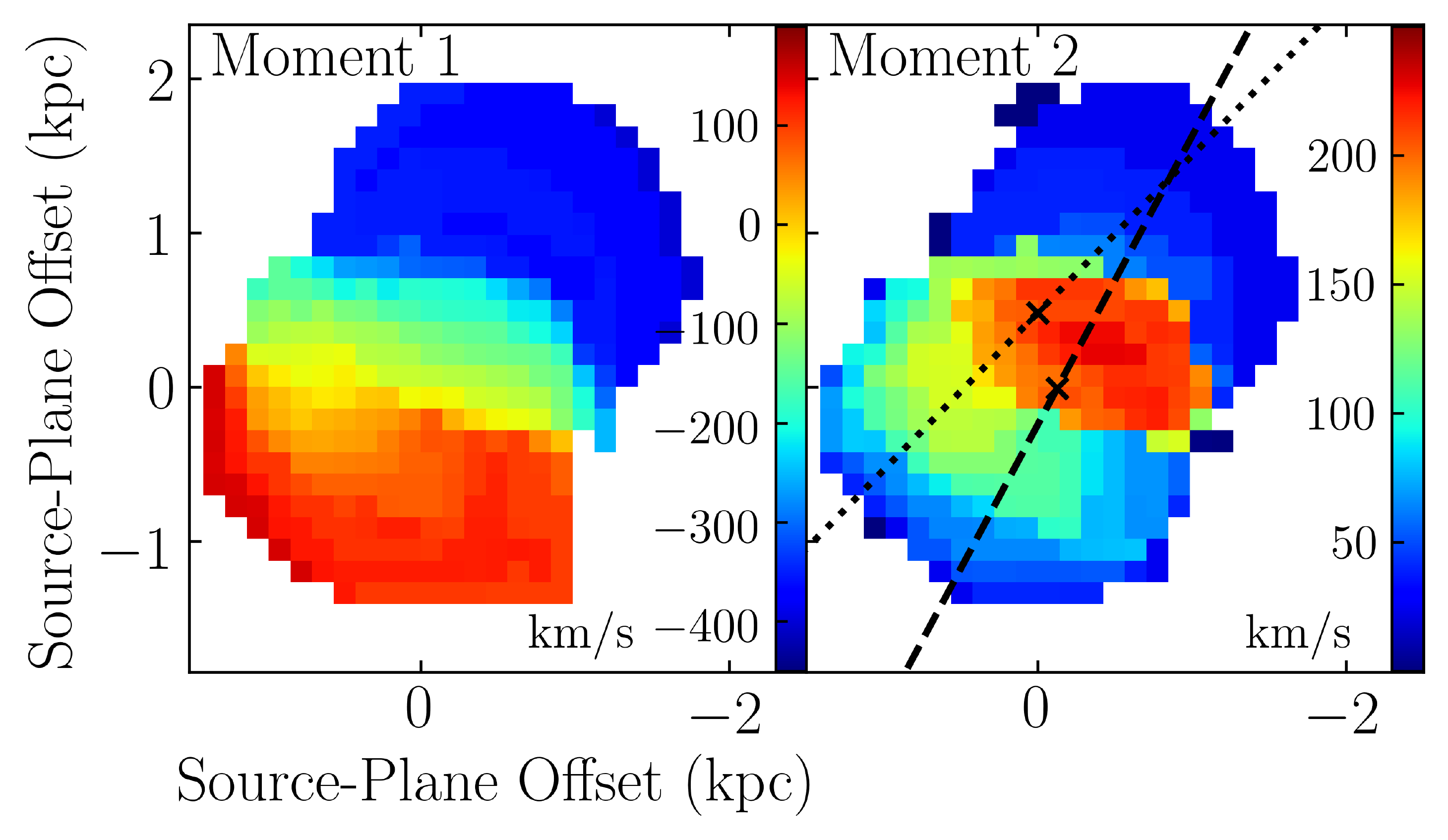}
\includegraphics[width=0.45\textwidth]{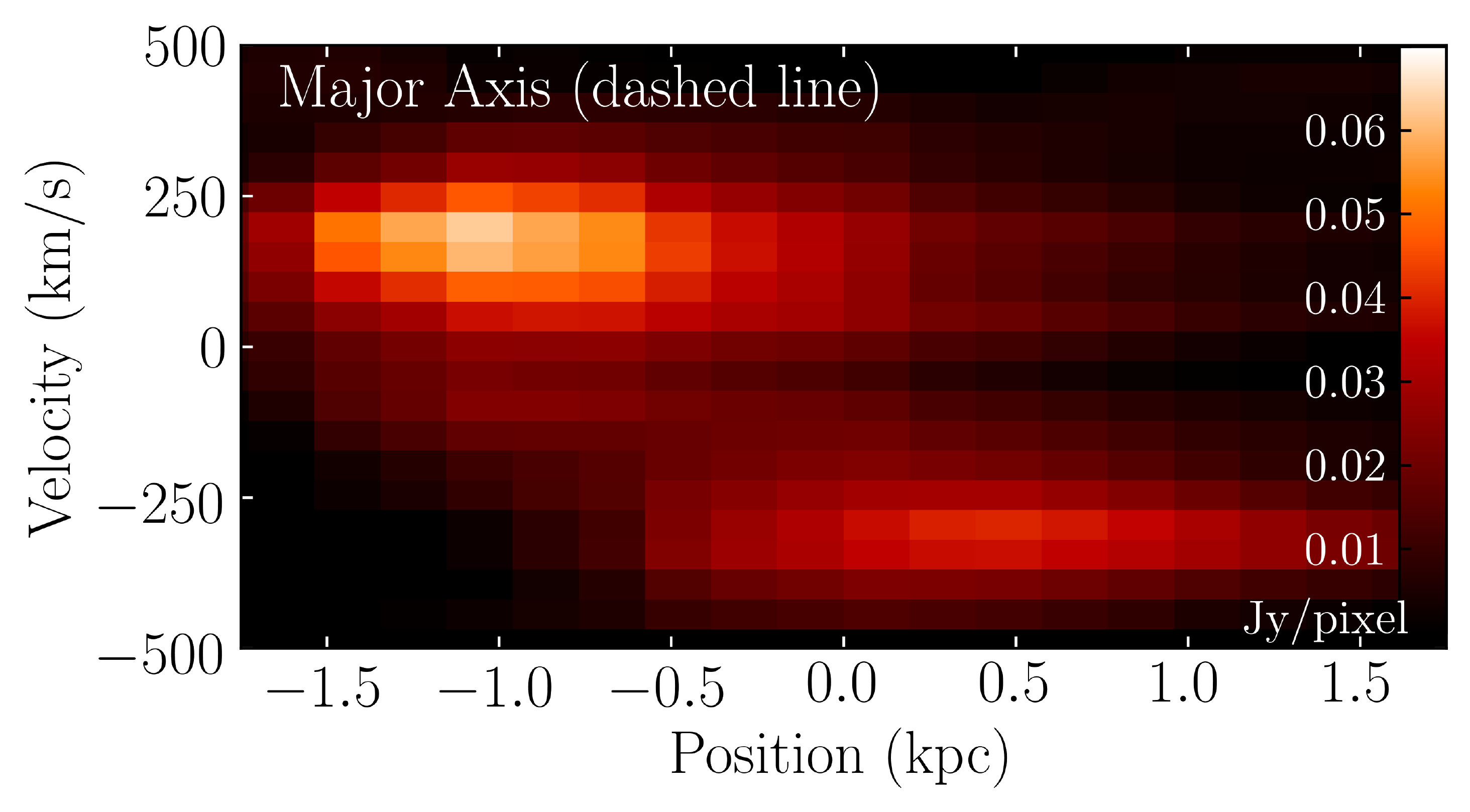}
\includegraphics[width=0.45\textwidth]{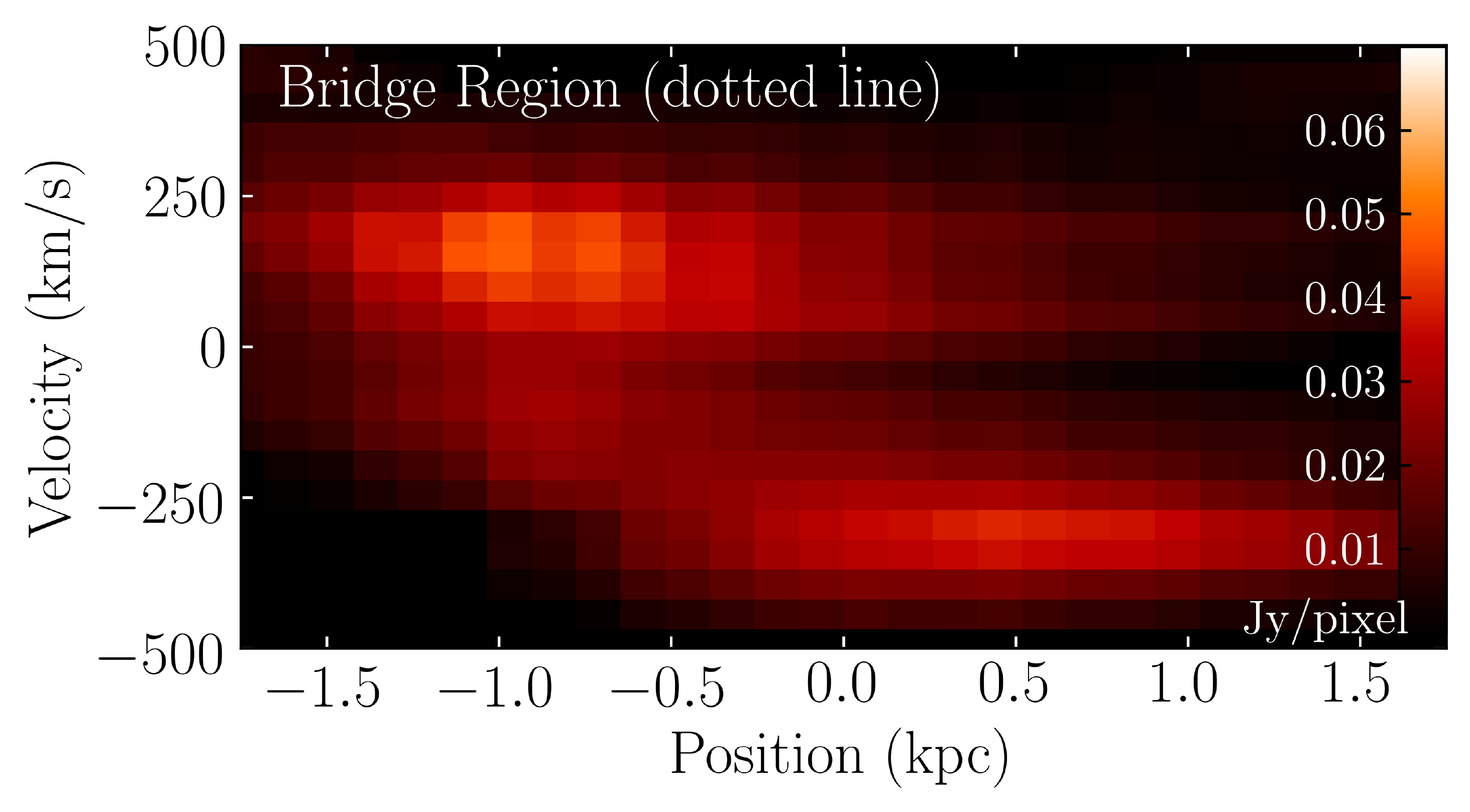}
\caption{Top:  Moments 1 and 2 of [CII] in SPT0346-52.  Middle:  Position-velocity diagram of the major axis of SPT0346-52.  The velocities were extracted along the dashed line shown in the moment 2 map. There are two spatially and kinematically components.  Bottom:  Position-velocity diagram of the bridge connecting the two components of SPT0346-52.  The velocities were extracted along the dotted line shown in the moment 2 map, with positions $0\ \rm{kpc}$ marked by the black x's.
The units on the colorbars in the position-velocity diagrams are Jy/pixel. \label{fig:pv}}
\end{center}
\end{figure}

In addition to the \cii\ emission map shown in Figure~\ref{fig:source}, we calculate moment 1 (intensity-weighted average velocity, shown in the top left panel of Figure~\ref{fig:pv}) and moment 2 (intensity-weighted velocity dispersion, top right panel of Figure~\ref{fig:pv}) of the reconstructed line.
The velocity dispersions in the center of the system reach very high values ($\sigma > 200\ \rm{km\ s^{-1}}$).
Extracting the velocities along the major axis of SPT0346-52 (dashed line in Figure~\ref{fig:pv}, top right panel) reveals two spatially distinct velocity components.
This is shown in the position-velocity diagram in the middle panel of Figure~\ref{fig:pv}.

\begin{figure}
\begin{center}
\includegraphics[width=0.4\textwidth]{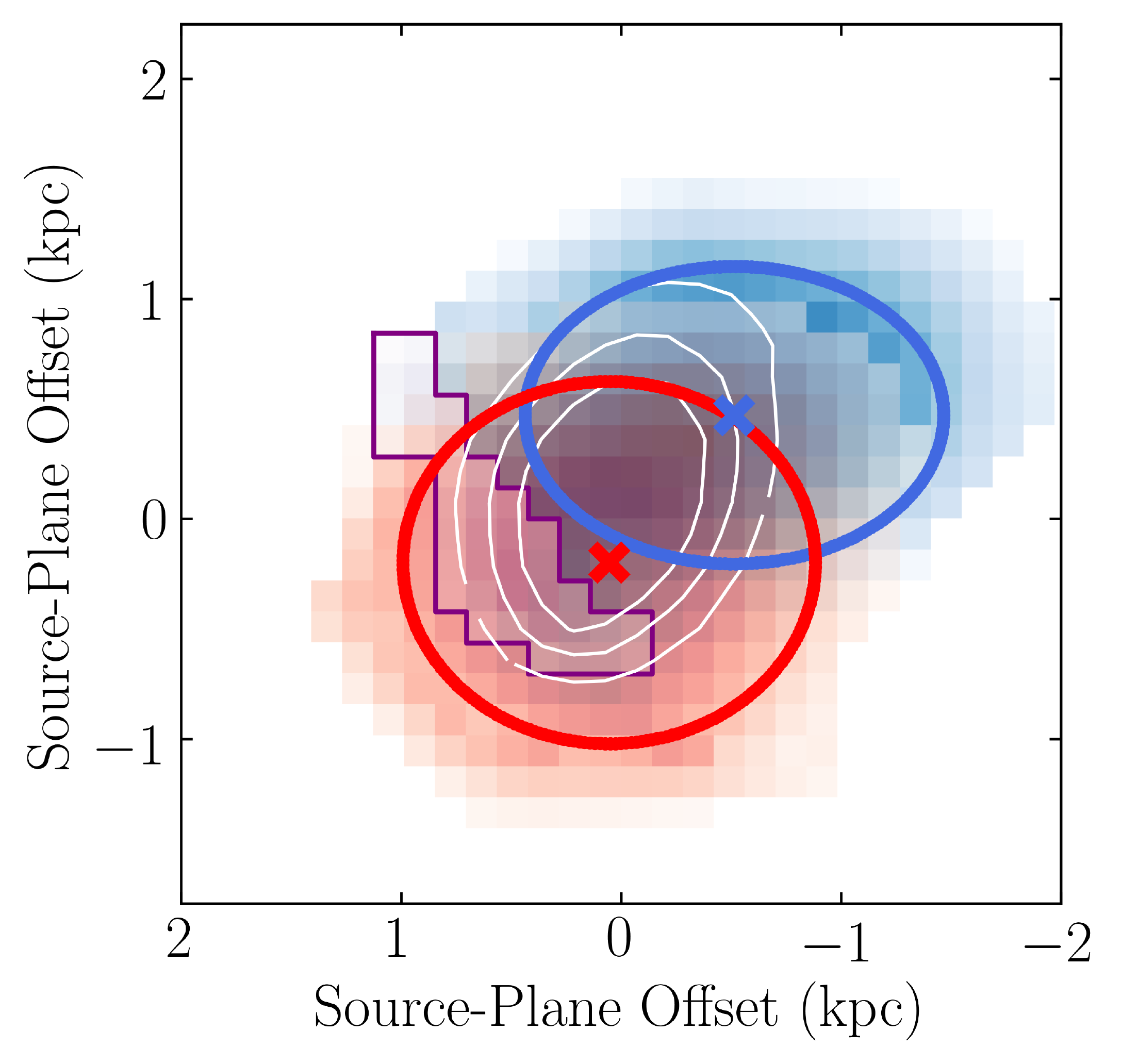}
\caption{Separation of the two components in SPT0346-52.
The blue and red ellipses outline the FWHM size of the blue and red components, and the centers are marked with an x.  
The amplitudes of the red and blue components of the Gaussian fits are shown as the red and blue images.  The darker region in the center shows where the components overlap.
The purple line outlines the pixels containing the "bridge" connecting the two components.  Contours of the continuum emission are overlaid in white for reference.\label{fig:ell}}
\end{center}
\end{figure}

In order to separate the two velocity components seen in Figure~\ref{fig:pv}, we fit the spectrum of each pixel with two Gaussian components. 
Each Gaussian is assigned to the appropriate galaxy component based on its velocity.
The shape of the velocity-integrated \cii\ emission in each of the two spatial components is fitted with an elliptical gaussian distribution. The centers and elliptical full-width half-maximum shapes of these components are shown in Figure~\ref{fig:ell}. The spatial distribution of the gas bridge, outlined in purple in Figure~\ref{fig:ell}, is determined by selecting pixels with emission at velocities intermediate to the two main galaxy components.

\section{Discussion}
\label{sec:disc}

\subsection{Merging Galaxies}

Estimates of the fraction of DSFGs that have multiple components or are merging are varied.
For example, from continuum emission only \cite{spilker2016} found only $13\%$ of lensed DSFGs from the SPT sample showed strong evidence of having multiple components, while \cite{bussmann2015} found $69\%$ of DSFGs with multiple components.
If only the intensity-weighted velocity (moment 1) map (see Figure~\ref{fig:pv}) were considered when studying the kinematics of SPT0346-52, this system could appear to be a symmetric, rotating disk.
However, only $\sim 40\%-80\%$ of merging systems show asymmetric kinematics in their star-forming gas \citep{hung2016}, so symmetric gas kinematics is not a definitive way to determine that a galaxy is not a merging system.

The lensing reconstruction of SPT0346-52 reveals two separated components (see Figure~\ref{fig:pv}).
The centers of these components are separated by $\sim 1\ \rm{kpc}$ and $\sim 500\ \rm{km s^{-1}}$.
There is a significant decrease in emission at velocities between the center velocities of the two components, as shown in Figure~\ref{fig:srcspec}.
Both of these components are larger than the effective resolution in this region of the reconstructed source plane, and they are separated by $\sim 2-3$ resolution elements.
Therefore, these two components are more likely to represent two separate structures, rather than a barely-resolved rotating disk.

The two components overlap in the middle of the system, near the peak of the continuum emission.
This region of overlap has a more complex velocity structure and higher velocity dispersions (Figure~\ref{fig:pv}, top panel).
Because the overlap region is where there is the most dust continuum emission, the star formation is likely occurring most intensely in that region, as has been observed in other merging systems such as The Antennae Galaxies \citep{mirabel1998,karl2010}.
\cite{teyssier2010} also found that merger-induced star formation is relatively concentrated near the center of merging systems in their hydrodynamic simulations.

In addition to the red and blue components, a position-velocity slice through the more complex velocity structure reveals a bridge of gas connecting the two components. The extraction line is indicated by the dotted line in the moment 2 map in Figure~\ref{fig:pv}, and the position-velocity diagram through this slice is shown in the bottom panel of Figure~\ref{fig:pv}.
The location of this bridge feature is also indicated by the purple contour in Figure~\ref{fig:ell}.
This structure resembles simulated tidal tails and observed tidal tails, such as in Arp105 \citep{bournaud2004}, as well as the south tail in the The Antennae \citep[NGC 4038/9,][]{gordon2001} in position-velocity diagrams.
\cite{decarli2017} also found \cii\ emission connecting a quasar host galaxy, PJ308-21, and a companion galaxy, though on much larger scales ($25\ \rm{kpc}$ and $1000\ \rm{km\ s^{-1}}$) than what is observed in SPT0346-52.

Both components of SPT0346-52 have large velocity dispersions ($> 200\ \rm{km\ s^{-1}}$, determined by the Gaussian fits to the spectra in each pixel). These large turbulent motions can help stabilize disks against gravitational fragmentation \citep[e.g.,][]{westmoquette2012,rangwala2015}, see Section \ref{sec:q}.

Several other merging DSFGs have been observed.
For example, \cite{neri2014} observed \cii\ emission in HDF850.1 ($z=5.185$) and found two components, one red-shifted and one blue-shifted, and separated by $2\ \rm{kpc}$ with radii $\sim 1\ \rm{kpc}$.
These components are similar in size to the components observed in SPT0346-52.
\cite{neri2014} explored the idea that HDF 850.1 was a rotating disk, but
concluded that they observed a merger-driven starburst.
\cite{rawle2014} also observed a late-stage merging DSFG at $z>5$ (HLS0918) with up to four components separated by $<4\ \rm{kpc}$.
\cite{engel2010} concluded using CO observations that most bright DSFGs with $\rm{L_{IR}} > 5\times 10^{12}\ \rm{\lsol}$ are major mergers.
This is consistent with the conclusion drawn from studies of stellar structures \citep[e.g.,][]{chen2015}.
At $z\sim 7$, \cite{hashimoto2018} concluded that the Lyman-break galaxy B14-65666 was a merger-induced starburst galaxy based on the velocity gradient in the \cii\ line and a two-component spectrum, whose spatial positions are consistent with two \cii\ knots and UV emission peaks.
Cosmological hydrodynamic galaxy formation simulations by \cite{narayanan2015} have shown that many DSFGs have multiple components, though the intense star formation may be driven by stellar feedback rather than major mergers.

Mergers can trigger intense star formation activity without producing an obvious AGN in DSFGs \citep{wang2013x}.
Though many ULIRGs, which have similar $\rm{L_{IR}}$ as DSFGs and enhanced star formation, have AGN activity that heats the dust and causes their high values of $\rm{L_{FIR}}$, \cite{younger2009} found that star formation alone can produce warm IR colors and produce UV radiation that is reradiated by hot dust.
About $63\%$ of luminous infrared galaxies (LIRGs) have multiple components \citep{haan2011,engel2011}.
Merging ULIRGs, which have higher FIR luminosities than LIRGs and have FIR luminosities more similar to that of SPT0346-52, have small nuclear separations (average $1.2\ \rm{kpc}$) and are in later merging systems \citep{haan2011}.
Similarly, SPT0346-52 could be a late-stage merger.

\cite{pavesi2018} recently observed \cii\ in a DSFG similar to SPT0346-52 at $z=5.667$, COSMOS \mbox{(FIR-)}Red Line Emitter (CRLE), with $\rm{SFR} = 1500\ \rm{\msol\ yr^{-1}}$ and a diameter of $2.7\ \rm{kpc}$.
They determined that CRLE is an intermediate stage merger.
CRLE has a gas depletion time scale of $45\ \rm{Myr}$.
For SPT0346-52, we calculate a gas depletion timescale of $31 \pm 10\ \rm{Myr}$ by dividing the gas mass from \cite{spilker2015} by the star formation rate from \cite{ma2015}, similar to the depletion timescale calculated by \cite{aravena2016} for this system.

An alternative explanation for the kinematic morphology in SPT0346-52 is that it is a rotating galaxy with a clumpy gas disk.
Clumpy, rotating disks have been observed in DSFGs \citep[e.g.,][]{hodge2012,iono2016,dannerbauer2017,tadaki2018}.
However, \cite{hodge2016} searched for $\sim 1\ \rm{kpc}$ clumps (comparable to the sizes of the clumps in GN20 and the components in SPT0346-52) in luminous DSFGs and found no significant evidence for clumping in most cases.
\cite{gullberg2018} looked at \cii\ in four $z\sim 4.5$ DSFGs.  They found three that showed a smooth morphology, while the fourth could be a clumpy disk, though they cannot rule out the possibility of it being a smooth disk.
The data explored by \cite{gullberg2018} and \cite{hodge2016} did not have enough signal-to-noise to definitively show that the observed clumps were real, rather than noise fluctuations.
The data presented in this work have a higher signal-to-noise ratio, allowing a more confident classification of this system as a merger rather than a clumpy, rotating disk.

\subsection{Stability of Components}
\label{sec:q}

The Toomre Q parameter describes the stability of rotating disk against gravitational collapse.  It is given by
\begin{equation}
Q = c_s \kappa \frac{1}{\pi G \Sigma_{\rm gas}},
\end{equation}
where $c_s$ is the sound speed, $\kappa$ is the epicyclic frequency, and $\Sigma_{\rm gas}$ is the gas surface density \citep{toomre1964}.
In a system dominated by turbulent pressure, rather than thermal pressure, this becomes
\begin{equation}
Q = \sqrt{c_s^2+\sigma_T^2} \kappa \frac{1}{\pi G \Sigma_{\rm gas}},
\end{equation}
where $\sigma_T$ is the turbulent velocity dispersion.
In the limit of high turbulence, $\sqrt{c_s^2+\sigma_T^2} \sim \sigma_T$ \citep{hayward2017}.
The gas is stable against gravitational collapse if $Q>1$ and unstable for $Q<1$, though observations of galaxies and simulations of thick disks place this threshold at $Q\sim 0.7$ \citep{kennicutt1989,kim2007}.

The sound speed and turbulent linewidth are difficult to measure directly, so the gas velocity dispersion, $\sigma_r$ is often used instead. 
In cosmological simulations from the FIRE (Feedback In Realistic Environments) suite,
\cite{su2017} found that stellar feedback, which would be an important factor in a rapidly star-forming system like SPT0346-52, increases the turbulent velocity dispersion by a factor of 2-3. 
Using the velocity dispersion instead of the sound speed and true turbulent velocity dispersion likely provides an upper limit to Q \citep{prieto2016}. Because the observed velocity dispersion ($\sigma_r$) can include ordered motion such as rotation or outflows it tends to overestimate $\sigma_T$.\citep{su2017}.

The epicyclic frequency, $\kappa$, is $a\Omega$, where $\Omega$ is the rotational frequency and $a$ is a constant.  For a flat rotation curve, $a=1$.  In general, $1<a<2$.
\cite{swinbank2015} and \cite{oteo2016} used an intermediary value of $a=\sqrt{3}$; we use the same substitution here.
The rotational frequency can be described as $\Omega = v_r/r$.
Then, $\kappa =a\Omega \approx \sqrt{3} v_r/r$.

With the above substitutions, we calculate the Toomre Q stability parameter using
\begin{equation}
Q \approx \sigma_r \frac{\sqrt{3} v_{r}}{r} \frac{1}{\pi G \Sigma_{\rm gas}}. \label{eq:Q}
\end{equation}
While we do not assume that the components of SPT0346-52 are disks, past spatially resolved calculations of Q have found $Q<1$ locally where there are
star-forming regions and giant molecular clouds in other systems, even when the global disk has $Q>1$ \citep[i.e.,][]{fisher2017,genzel2011,martig2009}.
The Q parameter can therefore be used to find local instabilities independent of the global stability/instability of a system.

To calculate the gas surface density, $\Sigma_{\rm gas}$, we assume the \cii\ emission traces the gas.  The total gas mass, $M_g = 1.5\times 10^{11}\ \rm{\msol}$, taken from \cite{spilker2015}, is divided among the pixels according to their \cii\ luminosity.  To convert to surface density, the gas mass in each pixel is divided by the area of the pixel.

The surface density is then given by
\begin{equation}
\Sigma_{\rm gas,i} \approx M_g \frac{S_i}{\sum_i S_i} \frac{1}{A_i},
\end{equation}
where $S_i$ is the integral of the Gaussian component in each pixel for each component from Section \ref{sec:kin}, $\sum_i S_i$ is the total \cii\ flux density, and $A_i$ is the area of a pixel.

The value of $\sigma_r$ used in Equation~\ref{eq:Q} is the standard deviation determined in the Gaussian line fitting described in Section \ref{sec:kin}.  
To calculate $v_r$, we first created velocity fields for the two spatial/velocity components using the mean velocity determined by the two-Gaussian line fitting described in Section \ref{sec:kin}.
These velocity fields were then fit using the 2D tilted ring modeling in 3D-Barolo \citep{bbarolo}.
These model velocity fields are used as the values of $v_r$ throughout both components.
The position, $r$, is defined relative to the center of each component, indicated by red crosses in Figure~\ref{fig:ell}.

\begin{figure}
\begin{center}
\includegraphics[width=0.45\textwidth]{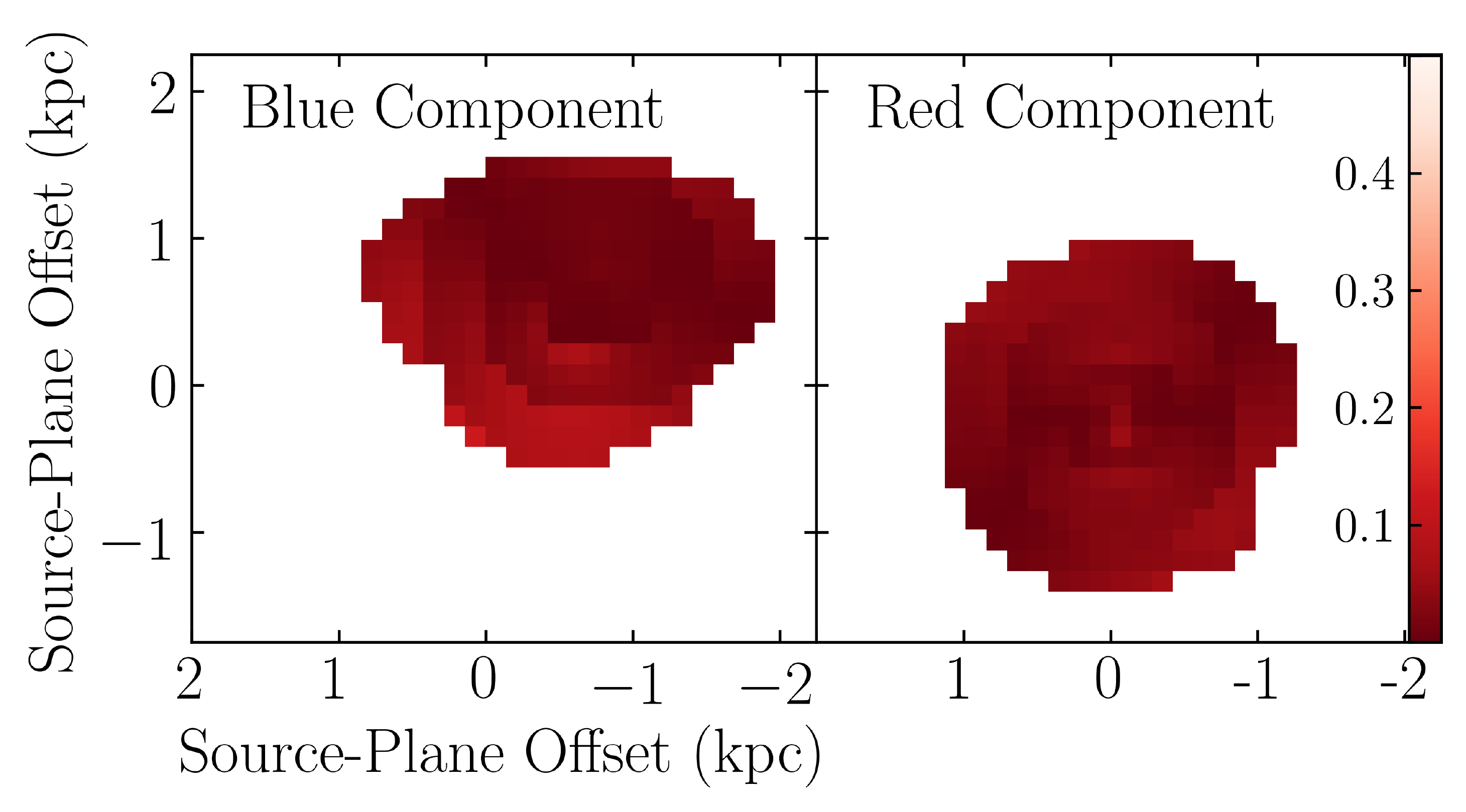}
\caption{Maps of Toomre Q disk stability parameter. Left:  blue component. The blue component pixels have \cii\ intensity-weighted mean $\bar{Q}=0.03$ and maximum $Q_{\rm max} = 0.13$. Right:  red component.  The red component pixels have $\bar{Q}=0.02$ and $Q_{\rm max} = 0.06$.  $Q \ll 1$ throughout the galaxy indicates that it is unstable against gravitational collapse, consistent with rapid star formation throughout both disks.  \label{fig:q}}
\end{center}
\end{figure}

A map of the Toomre Q stability parameter is shown in Figure~\ref{fig:q}.
The individual pixels in the blue component have a \cii\ intensity-weighted mean of $\bar{Q}=0.03$ and a maximum value of $Q_{\rm max}=0.13$.
The individual pixels in the red component have  $\bar{Q}=0.02$ and $Q_{\rm max}=0.06$.
All values of Q are well less than one, indicating that the the system (separated into individual components) is unstable to gravitational collapse.
As mentioned above, using $\sigma_r$ instead of $\sigma_T$ or $c_s$ gives the upper limit for Q.
Thus, the result that $Q\ll 1$ everywhere and the disks are gravitationally unstable does not depend on this substitution.

We also calculated values of Q that would be measured for SPT0346-52 if we could not spatially resolve its structure, as is typical of high-redshift galaxies observed to date. We also consider the unresolved estimates for the red and blue components alone.  These values, along with the maximum rotational velocity, $V_{\rm max}$, the mean velocity dispersion, $\bar{\sigma}$, and the radius of each component, $R$, are given in Table \ref{table:q}.
The values of Q are low compared to previous studies of DSFGs.
For example, \cite{oteo2016} calculated $Q\sim 0.22$ and $Q\sim 0.35$ for an interacting pair of DSFGs at $z = 4.425$, and \cite{swinbank2015} found $Q\sim 0.3$ in SDP.81.
\cite{debreuck2014} found a higher average in ALESS 73.1, a z=4.76 DSFG, with average $Q=0.58$, though with $Q<1$ at all radii.
In Arp220, $Q<1$ only in the inner part of the disk, where the most intense star formation is occurring \citep{scoville1997}.
The values of Q in SPT0346-52 are consistent with studies of star-forming galaxies, where giant star-forming clumps and local overdensities were found to be unstable against fragmentation \citep{genzel2011,westmoquette2012,martig2009}.
Where $Q>1$ in Seyfert galaxies and ULIRGs, the disks cannot fragment and form stars \citep{sani2012,tacconi1999}.
The low values of Q throughout SPT0346-52 indicate that the components are very unstable against collapse, which is fully consistent with the observed high star formation rate.

\begin{deluxetable}{rccccc}
\tablecaption{Kinematic Values of the Components of SPT0346-52}
\tablenum{3}
\label{table:q}
\tablehead{\colhead{Source} & \colhead{$V_{\rm c}$} & \colhead{$V_{\rm max}$} & \colhead{$\bar{\sigma}$} & \colhead{$R$} & \colhead{$Q$}}\\
\startdata
 & ($\rm {km\ s^{-1}}$) & ($\rm{km\ s^{-1}}$) & ($\rm{km\ s^{-1}}$) & kpc &  \\
\hline
Both & & $263$ & $307$ & $1.6$ & $0.30$ \\
Blue & $-309$ & $95$ & $301$ & $0.95$ & $0.11$ \\
Red & $158$ & $30$ & $313$ & $0.94$ & $0.05$ 
\enddata
\end{deluxetable}

\subsubsection{The Future of SPT0346-52}

While mergers can trigger the onset of an AGN \citep[e.g.,][]{wang2013x}, SPT0346-52 has negligible AGN activity \citep{ma2016}.
However, many DSFGs and merging systems do have AGN \citep[e.g.,][]{rawle2014,carniani2013,westmoquette2012,engel2011,younger2009,younger2008}.
At $z=3.351$, \cite{marsan2015} found an AGN in an ultra-massive and compact galaxy at $z=3.35$ whose stars formed in an intense starburst $300-500\ \rm{Myr}$ prior.
It is possible SPT0346-52 currently has an AGN that is so heavily obscured that X-ray emission is not visible. 
SPT0346-52 may also host an AGN in the future.

DSFGs are thought to evolve to form the red sequence by $z=2$.
The stars in this red sequence would form in an intense, short, dissipative burst of star formation at $z > 4$ within a compact, $r_e\approx 1\ \rm{kpc}$, region \citep{kriek2008}.
This effective radius is similar to that of SPT0346-52.
The models by \cite{narayanan2015} suggest that by $z\sim 0$ DSFGs (like SPT0346-52) will reside in massive dark matter (DM) halos with $M_{\rm{DM}} \approx 10^{14}\ \rm{\msol}$, though not all of the intense star formation is driven by major mergers.
These studies are in agreement with that of \cite{cattaneo2013}, who found that most $2<z<4$ star-forming galaxies with $M_* > 10^{10}\ \rm{\msol}$ evolve into the most massive galaxies on the red sequence and had a phase of intense star formation at $z>2$.
Similarly, angular clustering analyses of $z>2$ blank-field DSFGs have suggested that DSFGs evolve into present day halos with masses of $10^{13}-10^{14}\ \rm{\msol}$ \citep[e.g.,][]{chen2016,wilkinson2017}.
\cite{oteo2016} observed a pair of interacting DSFGs at $z=4.425$.
The system observed by \cite{oteo2016} is at an earlier merger stage than SPT0346-52.
They concluded that this system is likely the progenitor of a massive, red, elliptical galaxy.
At $z=2.3$, \cite{fu2013} studied two interacting massive starburst galaxies, separated by $19\ \rm{kpc}$ and connected by a tidal tail or bridge.
They similarly conclude that this system will deplete its gas reservoir in $200\ \rm{Myr}$ and merge to form an elliptical galaxy with $M_* \sim 4\times 10^{11}\ \rm{\msol}$.

SPT0346-52 is currently undergoing a phase of intense star formation.
It may deplete its gas reservoir in $\sim 30\ \rm{Myr}$ \citep{aravena2016,spilker2015,narayanan2015,fu2013} and evolve into a red sequence galaxy.

\section{Summary and Conclusions}
\label{sec:conc}

In this paper, we presented a pixellated lensing reconstruction of high-resolution \cii\ emission observed with ALMA towards the $z=5.7$ dusty star-forming galaxy SPT0346-52.
With this reconstruction, we mapped the integrated \cii\ emission and dust continuum at rest-frame $158\ \rm{\mu m}$ in the (unlensed) source plane.
We spatially resolved the $\rm{L_{\cii}/L_{FIR}}$ ratio in SPT0346-52 and showed that the $\rm{L_{\cii}/L_{FIR}}$ vs $\Sigma_{\rm FIR}$ relation continues at smaller spatial scales.

We also obtained source-plane velocity information on SPT0346-52, including a demagnified spectrum and moment maps.
The reconstruction revealed two spatially and kinematically separated components, one red-shifted and one blue-shifted relative to the \cii\ rest frequency.
These components are connected by a bridge of gas.
Each individual component is extremely unstable, with the Toomre Q stability parameter $Q \ll 1$ throughout both components.

These components are in the process of merging.
This merger is likely driving the intense star formation observed in SPT0346-52.
SPT0346-52 may have an AGN in its future and evolve into a massive red sequence galaxy.

\acknowledgments

K.C.L. and D.P.M. acknowledge support
from the U.S. National Science Foundation (NSF) under awards 
AST-1715213 and AST-1312950 and K.C.L. through award SOSPA4-007 from the
National Radio Astronomy Observatory (NRAO). 
J.V. acknowledges support from NSF award AST-1716127.
This material has made use of the El
Gato (supported by the NSF award MRI-1228509)
high-performance computer.
This paper makes use of the following ALMA data: ADS/JAO.ALMA\#2013.1.01231.S.
ALMA is a partnership of ESO (representing its member states), NSF (USA) and NINS (Japan), together with NRC (Canada), MOST and ASIAA (Taiwan), and KASI (Republic of Korea), in cooperation with the Republic of Chile. The Joint ALMA Observatory is operated by ESO, AUI/NRAO and NAOJ. The
SPT is supported by the National Science Foundation through
grant PLR-1248097, with partial support through PHY1125897,
the Kavli Foundation and the Gordon and Betty
Moore Foundation grant GBMF 947. 
The Flatiron Institute is supported by the Simons Foundation.
This research has made use of NASA’s Astrophysics Data System.

\bibliographystyle{apj}
\bibliography{dsfgs}

\clearpage

\appendix

\section{Using Visibility Data to Create the \cii\ Spectrum}
\label{app:spec}

SPT0346-52 is an extended source with an irregular structure due to gravitational lensing.
Therefore, there is no optimal aperture to contain all of the emission.
When imaging these data, one has to make assumptions about the structure of the source.
Different weightings of the visibilities emphasize different aspects of the galaxy's structure (i.e., faint emission or small structures) and suppress some of the information inherently available from the visibilities.
In contrast, the observed complex visibilities contain all of the spectral line information.

We therefore obtain a spectrum of the observed \cii\ emission from the observed complex visibilities.
The flux density in a given channel, $F_v$, is determined by
\begin{equation}
F_v = \frac{\sum_i \frac{\widetilde{v}_{v,i}}{\widetilde{m}_i} \vert \widetilde{m}_i\vert ^2}{\sum_i \vert \widetilde{m}_i\vert ^2},
\end{equation}
where
$\widetilde{v}_{v,i}$ is the complex line data visibility and
$\widetilde{m}_i$ is the complex model visibility for the integrated \cii\ line.
Dividing data by the model in the numerator removes the spatial structure, transforming the observed visibilities to a point source. 
The data visibilities are then weighted by the amplitude of the model visibilities in the sum, 
which emphasizes the visibilities where line emission is expected, and minimizes the weight of the visibilities with little-to-no sensitivity to the emission structure.
We use a model of the complex visibilities for the integrated \cii\ line created using a pixellated gravitational lensing reconstruction described further in Section \ref{sec:ripples}.

The error on the flux density in each channel, $\sigma$, is the quadrature sum of the weights such that
\begin{equation}
\sigma^2 = \left( \sum_i \vert \widetilde{m}_i \vert^2 \right)^{-1}.
\end{equation}

The resulting spectrum is shown in Figure \ref{fig:spec}.

\end{document}